\documentclass{article}

\usepackage{arxiv}

\usepackage[utf8]{inputenc} 
\usepackage[T1]{fontenc}    
\usepackage{hyperref}       
\usepackage{url}            
\usepackage{booktabs}       
\usepackage{amsfonts}       
\usepackage{nicefrac}       
\usepackage{lipsum,enumerate,bm,mathrsfs,yfonts}
\usepackage{graphicx,amsmath,multicol,multirow}
\usepackage{url}
\usepackage{amsthm}
\usepackage[style=apa, backend=biber, natbib=true]{biblatex}
\usepackage{adjustbox,xr,comment} 
\usepackage{pdfpages}

\usepackage{amsmath}
\usepackage{amssymb}
\usepackage{amsthm,tikz}
\usetikzlibrary{matrix,positioning,calc}
\usepackage{colortbl,caption,xr}
\usepackage{adjustbox}  
\usepackage{graphicx}
\usepackage{lscape}
\usepackage{rotating}
\usepackage{graphicx}
\usepackage{float}
\usepackage{url}
\usepackage{subfigure}
\usepackage{dcolumn}
\usepackage{multirow}
\usepackage{verbatim,booktabs, array,arydshln}
\usepackage{color}
\usepackage{setspace}
\usepackage{bbm}
\usepackage{booktabs}
\usepackage{hhline}
\usepackage{bm}
\usepackage{dsfont,import}
\usepackage{enumerate}
\usepackage{threeparttable}
\usepackage{array}
\usepackage{caption}

\usepackage{moreverb}
\usepackage{bm}
\usepackage{hyperref,rotating,graphicx}
\usepackage{xr}
\usepackage{verbatim}
\usepackage{url}
\theoremstyle{definition}

\usepackage{tcolorbox}
\usepackage{amsmath}
\usepackage{geometry}
\usepackage{fancyhdr}
\usepackage{placeins}
\usepackage{xcolor}
\usepackage{booktabs,tabularx,array}
\newcolumntype{Y}{>{\centering\arraybackslash}X}

\theoremstyle{definition}

\theoremstyle{remark}

\makeatletter

\title{Cluster randomized crossover trials with very few clusters but multiple periods: which analyses for continuous outcomes should be used?}

\author{
 Guangyu Tong$^{\dagger}$ \\
    Department of Internal Medicine \\
    Yale School of Medicine \\
    Department of Biostatistics, \\
    Center for Methods in Implementation and Prevention Science \\
    Yale School of Public Health \\
    New Haven, CT, USA\\
   \And
Qianzhe Sun$^{\dagger}$ \\
    Department of Biostatistics \\
    Yale School of Public Health \\
    New Haven, CT, USA\\
\And
Jessica Kasza\\
School of Public Health and Preventive Medicine\\
Monash University\\
Melbourne, Australia
\And
Andrew B. Forbes \\
School of Public Health and Preventive Medicine\\
Monash University\\
Melbourne, Australia
  \And
Monica Taljaard\\
Methodological and Implementation Research Program\\
The Ottawa Hospital Research Institute\\
School of Epidemiology and Public Health\\
University of Ottawa\\
Ottawa, ON, Canada
\And
Fan Li\\
Department of Biostatistics, \\
Center for Methods in Implementation and Prevention Science\\
    Yale School of Public Health \\
    New Haven, CT, USA\\
  \texttt{fan.f.li@yale.edu} \\[0.8em]
  $^{\dagger}$These authors contributed equally to this work.
}

\begin{document}
\maketitle
\begin{abstract}
Cluster randomized crossover (CRXO) trials are often used when individual randomization is impractical and the number of available clusters is limited. However, statistical analysis of CRXO trials is complex because of the need to account for complex correlation structures over time. It becomes especially challenging when very few clusters are used because standard modeling assumptions may lead to unstable variance estimates, poor confidence interval coverage, and inflated type I error. This study evaluates individual-level mixed-effects and fixed-effects models with and without a cluster-period random effect, cluster-period summary analysis using normal- or \(t\)-based inference, and two-period crossover-difference estimators. Using extensive simulation studies under both nested exchangeable and discrete time decay correlation structures, we compare model performance in terms of bias, root mean squared error, coverage probability, type I error, and convergence. Across scenarios, all models produced approximately unbiased treatment effect estimates, but their inferential performance differed substantially. Models that explicitly accounted for cluster-period heterogeneity generally provided the most reliable control of coverage and type I error, whereas simpler exchangeable models performed adequately only when the true correlation structure closely matched their assumptions. Cluster-period level analysis performance improved with increasing numbers of periods but was unreliable in the sparsest designs. Overall, the findings suggest that in CRXO trials with very few clusters, accurate modeling of cluster-period correlation is more important than the choice between fixed and random cluster intercepts, and that results from extremely sparse designs should be interpreted with caution.
\end{abstract}


\section{Background and Introduction}\label{sec:introduction}

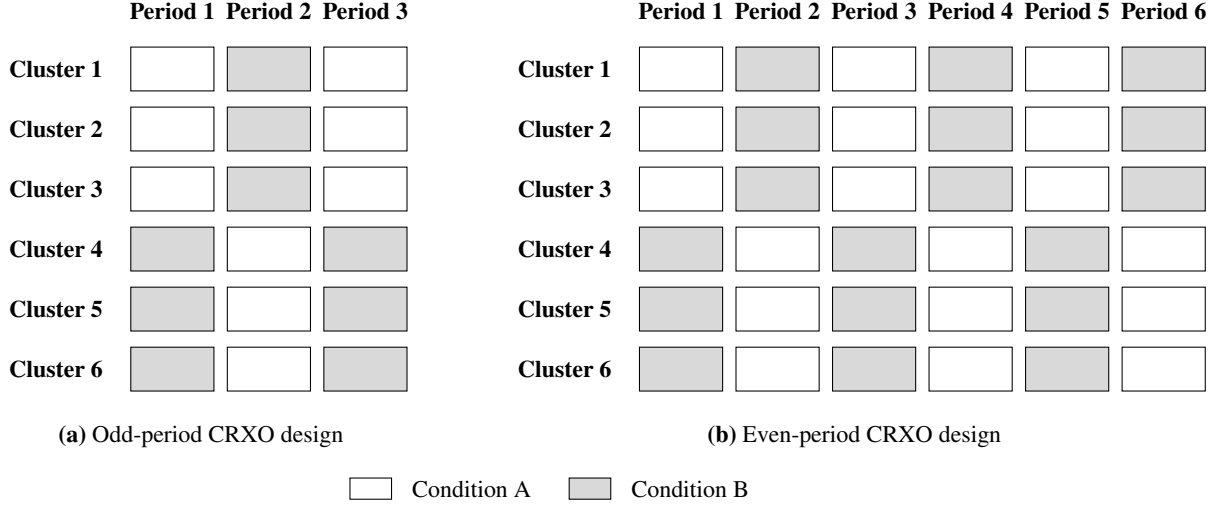
\begin{figure}[htbp]
    \centering

    \definecolor{conditionAcolor}{RGB}{255,255,255}
    \definecolor{conditionBcolor}{RGB}{217,217,217}

    \adjustbox{max width=\linewidth}{%
    \begin{tikzpicture}

        \matrix (odd) [
            matrix of nodes,
            matrix anchor=north west,
            ampersand replacement=\&,
            nodes in empty cells,
            nodes={
                draw=black,
                minimum width=1.10cm,
                minimum height=0.58cm,
                inner sep=0pt,
                anchor=center,
                font=\small
            },
            column sep=1.5mm,
            row sep=2mm,
            column 1/.style={
                nodes={
                    draw=none,
                    minimum width=1.65cm,
                    anchor=east,
                    font=\small
                }
            },
            row 1/.style={
                nodes={
                    draw=none,
                    minimum height=0.55cm,
                    font=\small
                }
            }
        ]{
            \&
            \textbf{Period 1}
            \& \textbf{Period 2}
            \& \textbf{Period 3} \\

            \textbf{Cluster 1}
            \& |[fill=conditionAcolor]|
            \& |[fill=conditionBcolor]|
            \& |[fill=conditionAcolor]| \\

            \textbf{Cluster 2}
            \& |[fill=conditionAcolor]|
            \& |[fill=conditionBcolor]|
            \& |[fill=conditionAcolor]| \\

            \textbf{Cluster 3}
            \& |[fill=conditionAcolor]|
            \& |[fill=conditionBcolor]|
            \& |[fill=conditionAcolor]| \\

            \textbf{Cluster 4}
            \& |[fill=conditionBcolor]|
            \& |[fill=conditionAcolor]|
            \& |[fill=conditionBcolor]| \\

            \textbf{Cluster 5}
            \& |[fill=conditionBcolor]|
            \& |[fill=conditionAcolor]|
            \& |[fill=conditionBcolor]| \\

            \textbf{Cluster 6}
            \& |[fill=conditionBcolor]|
            \& |[fill=conditionAcolor]|
            \& |[fill=conditionBcolor]| \\
        };

        \matrix (even) [
            matrix of nodes,
            matrix anchor=north west,
            right=1.0cm of odd.north east,
            ampersand replacement=\&,
            nodes in empty cells,
            nodes={
                draw=black,
                minimum width=1.10cm,
                minimum height=0.58cm,
                inner sep=0pt,
                anchor=center,
                font=\small
            },
            column sep=1.5mm,
            row sep=2mm,
            column 1/.style={
                nodes={
                    draw=none,
                    minimum width=1.65cm,
                    anchor=east,
                    font=\small
                }
            },
            row 1/.style={
                nodes={
                    draw=none,
                    minimum height=0.55cm,
                    font=\small
                }
            }
        ]{
            \&
            \textbf{Period 1}
            \& \textbf{Period 2}
            \& \textbf{Period 3}
            \& \textbf{Period 4}
            \& \textbf{Period 5}
            \& \textbf{Period 6} \\

            \textbf{Cluster 1}
            \& |[fill=conditionAcolor]|
            \& |[fill=conditionBcolor]|
            \& |[fill=conditionAcolor]|
            \& |[fill=conditionBcolor]|
            \& |[fill=conditionAcolor]|
            \& |[fill=conditionBcolor]| \\

            \textbf{Cluster 2}
            \& |[fill=conditionAcolor]|
            \& |[fill=conditionBcolor]|
            \& |[fill=conditionAcolor]|
            \& |[fill=conditionBcolor]|
            \& |[fill=conditionAcolor]|
            \& |[fill=conditionBcolor]| \\

            \textbf{Cluster 3}
            \& |[fill=conditionAcolor]|
            \& |[fill=conditionBcolor]|
            \& |[fill=conditionAcolor]|
            \& |[fill=conditionBcolor]|
            \& |[fill=conditionAcolor]|
            \& |[fill=conditionBcolor]| \\

            \textbf{Cluster 4}
            \& |[fill=conditionBcolor]|
            \& |[fill=conditionAcolor]|
            \& |[fill=conditionBcolor]|
            \& |[fill=conditionAcolor]|
            \& |[fill=conditionBcolor]|
            \& |[fill=conditionAcolor]| \\

            \textbf{Cluster 5}
            \& |[fill=conditionBcolor]|
            \& |[fill=conditionAcolor]|
            \& |[fill=conditionBcolor]|
            \& |[fill=conditionAcolor]|
            \& |[fill=conditionBcolor]|
            \& |[fill=conditionAcolor]| \\

            \textbf{Cluster 6}
            \& |[fill=conditionBcolor]|
            \& |[fill=conditionAcolor]|
            \& |[fill=conditionBcolor]|
            \& |[fill=conditionAcolor]|
            \& |[fill=conditionBcolor]|
            \& |[fill=conditionAcolor]| \\
        };

        \node[
            below=2mm of odd.south,
            anchor=north,
            font=\small
        ] (oddcaption)
        {\textbf{(a)} Odd-period CRXO design};

        \node[
            below=2mm of even.south,
            anchor=north,
            font=\small
        ] (evencaption)
        {\textbf{(b)} Even-period CRXO design};

        \coordinate (legendcenter) at
        ($(oddcaption.south)!0.5!(evencaption.south)+(0,-0.425cm)$);

        \draw[
            fill=conditionAcolor,
            draw=black
        ]
        ($(legendcenter)+(-2.35,-0.14)$)
        rectangle ++(0.55,0.28);

        \node[
            anchor=west,
            font=\small
        ]
        at ($(legendcenter)+(-1.65,0)$)
        {Condition A};

        \draw[
            fill=conditionBcolor,
            draw=black
        ]
        ($(legendcenter)+(0.55,-0.14)$)
        rectangle ++(0.55,0.28);

        \node[
            anchor=west,
            font=\small
        ]
        at ($(legendcenter)+(1.25,0)$)
        {Condition B};

    \end{tikzpicture}%
    }

    \caption{
    Illustration of balanced cluster randomized crossover designs.
    Panel (a) shows a three-period design in which clusters are equally
    allocated to the complementary alternating sequences \(ABA\) and
    \(BAB\).
    Panel (b) shows a six-period design in which clusters are equally
    allocated to the complementary alternating sequences \(ABABAB\) and
    \(BABABA\).
    }
    \label{fig:crxo_designs}
\end{figure}

Cluster‐randomized crossover (CRXO) trials are an important class of experimental design in basic science, clinical, social science and population health research, offering efficiency by allowing each cluster to serve as its own control.\citep{senn2002,jones2003,mills2009} In a CRXO design, intact groups ("clusters") such as hospitals, intensive care units (ICUs), schools, or clinical practices are randomized to a sequence of interventions over time, so that each cluster receives each intervention in one or more study periods. More specifically, the conventional design consists of two periods, with clusters randomized to one of two treatment sequences and receiving each study condition once. More recently, this framework has been extended to multiple-period designs, in which clusters cross between treatment conditions multiple times over more than two periods.\citep{hemming2020mpcc} In a systematic review of 91 published CRXO trials, Arnup et al.\citep{arnup2016appropriate} reported that 31\% of trials with reported period information included more than two periods, including 20\% with four or more periods, suggesting that multiple-period CRXO designs are not uncommon in practice. Although cohort CRXO designs, in which the same participants are followed across periods, are possible, repeated cross-sectional designs are more commonly used in practice, often with continuous recruitment such that different participants contribute outcomes in different periods. Representative balanced multiple-period CRXO designs with even and odd numbers of periods are illustrated in Figure~\ref{fig:crxo_designs}.

The CRXO design is especially attractive when individual randomization is infeasible because of contamination, logistics, or implementation constraints, when interventions can be switched back and forth across study periods,and when conventional parallel cluster-randomized trial would require more clusters than are practically available.\citep{hemming2020mpcc, senn2002, jones2003, mills2009} For example, in preclinical pharmacology, rodent studies often alternate treatments across the same cohort to evaluate drug effects with limited animal numbers.\citep{festing2002guidelines,festing1994reduction} In critical care, ICU “switch-back” trials have been used to evaluate system-level strategies, and in these settings, where the number of available ICUs may be limited, CRXO designs have been specifically recommended.\citep{Bellomo2013} One illustrative study alternated ICUs between selective digestive decontamination and standard care,\citep{desmet2009decontamination} while others have alternated between different ventilation protocols.\citep{girardis2016oxygen} By allowing each cluster to contribute observations under both intervention conditions, CRXO designs can recover some of the efficiency otherwise lost through cluster randomization and can reduce sensitivity to baseline differences between clusters.\citep{senn2002,chow2008, piantadosi2017, rietbergen2011crxo_design} These advantages have made CRXO and related multiple-period cluster crossover designs increasingly attractive in pragmatic and comparative-effectiveness research, particularly in hospital-based settings when interventions can be introduced and withdrawn quickly and outcomes are routinely collected.\citep{piantadosi2017, arnup2016crxo_methods, mckenzie2025consort_crxo}

At the same time, the statistical analysis of CRXO trials can be complicated by two key design features: the need to account for period effects and the presence of complex within-cluster correlation structures over time. Period effects arise when outcomes change systematically across study periods because of secular trends, seasonal variation, or changes in the underlying study population. Because treatment allocation also varies across periods, failure to adjust appropriately for these temporal changes may confound the estimated treatment effect or lead to invalid inference. The efficiency of CRXO design arises from the interplay of two correlations: the similarity among individuals within the same cluster-period and the similarity among individuals in the same cluster across different periods.\citep{grantham2019crxo_crossover} Arnup et al.\citep{arnup2017crxo_tutorial} describe these as the within-cluster within-period correlation and the within-cluster between-period correlation, showing that the gain from crossover depends critically on the latter being positive. When between-period correlation is high, within-cluster comparisons remove substantial cluster-level heterogeneity; when it is low, the crossover offers little advantage over a parallel cluster trial. In multiple-period settings, additional complexity arises because correlations may decay with increasing temporal separation, so that outcomes measured closer together in time are more alike than outcomes measured farther apart. This has motivated work on more flexible correlation structures, including continuous-time decay models, and on design questions such as how often clusters should cross over during a fixed study duration.\citep{grantham2019crxo_crossover, arnup2017crxo_tutorial} Recent work has shown that, under exchangeable or nested exchangeable within-cluster correlation structures, increasing the number of crossovers does not improve statistical power when the number and duration of study periods are fixed and the design is balanced.\citep{Tanvir2026}

Despite these methodological developments, practical guidance for the analysis of small CRXO trials remains limited. Precisely what should be considered a small number of clusters has not been established; here, we consider designs with between two to six clusters to be "small". In such settings, empirical type I error and confidence interval coverage may vary depending on the design, analysis model, and underlying correlation parameters. One central methodological question concerns whether the cluster (e.g., health care facilities, ICU) should be modeled as a fixed effect or as a random effect. Fixed-effects models provide robustness by explicitly controlling for each cluster, but they do not support generalization beyond the observed sample. Random-effects models, in contrast, allow extrapolation to a larger population but rely on distributional assumptions that may be fragile when the number of clusters is small.\citep{turner2007crxo, parienti2007cluster_crossover} For example, in a four-ICU crossover trial, a random-effects model may yield unstable variance estimates and inflated type I error, whereas a fixed-effects model may sacrifice generalizability but deliver more reliable inference in practice.

This tension is not unique to crossover trials. Evidence from other cluster-randomized designs, such as stepped-wedge trials, highlights the challenges of small-sample settings.\citep{ford2020fewclusters,thompson2018misspecified} Previous simulation studies demonstrate that when the number of clusters is small (e.g., six or fewer), random-effects models frequently produce inflated type I error rates, while fixed-effects models achieve error rates closer to nominal levels.\citep{lee2024fe_swd,turner2007crxo} Consistent with this, reviews of published stepped-wedge trials have found that a substantial proportion of studies employ fixed-effects specifications.\citep{tong2025small_swd_review,turner2007crxo} These findings suggest that in sparse cluster designs, researchers often turn to fixed effects as a pragmatic response to the limitations of random-effects approaches. Despite the methodological scrutiny devoted to stepped-wedge trials, small crossover trials have received comparatively little systematic evaluation, even though they are highly prevalent. In particular, whether the challenges observed in stepped-wedge trials extend to small crossover designs remains largely unknown.

Our study addresses this gap by conducting a comprehensive evaluation of modeling strategies for two- and multiple-period CRXO trials with very few clusters, focusing on continuous outcomes analyzed using linear models under a repeated cross-sectional sampling framework, in which different individuals are observed across study periods. In addition to comparing fixed- and random-effects models, we evaluate alternative cluster-period analyses, distinguishing between procedures that use normal critical values and those that use finite-sample \(t\)-critical values for inference. Our data-generating processes include both nested exchangeable and discrete-time decay correlation structures, with the latter allowing within-cluster correlations to diminish as the separation between periods increases. We also vary key parameters-including cluster-period size, ICC, cluster autocorrelation (CAC), and period effects-to reflect the diversity of real-world trials. Finally, we explicitly examine model convergence, an often-overlooked but practically important issue when estimating complex models in small samples. By systematically integrating these design and analytic considerations, our work moves beyond a simple binary comparison of fixed versus random effects. We aim to provide clear, evidence-based recommendations on best practices for analyzing small crossover trials, balancing statistical rigor, robustness, and practical feasibility. In doing so, this study seeks to equip applied researchers in animal studies and critical care medicine with tools to conduct more reliable analyses, thereby strengthening the validity of findings from these widely used but methodologically challenging designs.

\section{Statistical Methods for Design and Analysis of Crossover Cluster Randomized Trials} 
\label{sec:methods}

This section provides a review of common modeling options for CRXO trials. We set up the notation for a trial with $I$ clusters and $J$ periods as follows: $Y_{ijk}$ is the health outcome of the $k$th individual ($k = 1, \dots, n_{ij}$) in the $i$th cluster ($i = 1, \dots, I$) and $j$th period ($j = 1, \dots, J$). Let $\theta$ denote the intervention effect, and $X_{ij}$ be the treatment indicator for the $i$th cluster during the $j$th period, where $X_{ij} = 1$ indicates if a cluster-period is assigned to intervention, and $X_{ij} = 0$ otherwise. Period is represented by $\phi_j$. Under the fixed effect specification, $\phi_j$ is a dummy indicator for period $j$ except for $\phi_1$, which is always set to 0 for identification constraints. The individual-level residual error is denoted as $\varepsilon_{ijk}$ and is assumed to be independently and identically distributed with variance $\sigma_e^2$. This work focuses on two- and multiple-period CRXO designs with continuous outcomes under repeated cross-sectional sampling, two intervention conditions, and balanced treatment-sequence crossover schemes, where clusters are randomly and equally allocated to two dual treatment sequences so that treatment allocation is balanced within each period.

\subsection{Correlation structures and variance parameterization}
We first describe the correlation structures used to characterize the dependence induced by repeated observations within clusters over time. These parameterizations are used both to interpret the candidate analysis models and to define the data-generating mechanisms in Section~\ref{sec:simulation}.
To fix notation, we first introduce the variance-component representation underlying the nested exchangeable structure:
\begin{equation}
  Y_{ijk}
  = \beta_0 + \theta X_{ij}
    + \phi_j
    + u_i + \gamma_{ij} + \varepsilon_{ijk},
  \label{eq:M1}
\end{equation}
where $u_i \sim N(0,\sigma_u^2)$ is the random cluster intercept, $\gamma_{ij} \sim N(0,\sigma_\gamma^2)$ is the random cluster-period effect, and \(\varepsilon_{ijk}\sim N(0,\sigma_e^2)\) is the individual-level error term.

\subsubsection{Nested exchangeable structure}

Under the variance-component representation in Equation~\ref{eq:M1}, the within-period ICC between two distinct individuals $k\neq \ell$ from the same cluster-period is defined as $\rho_w=\operatorname{Corr}(Y_{ijk},Y_{ij\ell})$ and is given by $\rho_w = (\sigma_u^2+\sigma_\gamma^2) / (\sigma_u^2+\sigma_\gamma^2+\sigma_e^2)$. The between-period ICC, defined as the correlation between two individuals from the same cluster but in different periods, is $\rho_b
= \sigma_u^2 / (\sigma_u^2+\sigma_\gamma^2+\sigma_e^2)$.
We define the cluster autocorrelation coefficient (CAC) as
$\kappa = \rho_b / \rho_w = \sigma_u^2 / (\sigma_u^2+\sigma_\gamma^2)$.
Equivalently, for a given within-period ICC \(\rho_w\) and CAC \(\kappa\), the variance components can be written as
\[
\sigma_u^2
=
\kappa\frac{\rho_w}{1-\rho_w}\sigma_e^2,
\qquad
\sigma_\gamma^2
=
(1-\kappa)\frac{\rho_w}{1-\rho_w}\sigma_e^2.
\]
Thus, the between-period ICC is \(\rho_b=\kappa\rho_w\). When \(\kappa=1\), \(\sigma_\gamma^2=0\), and the structure reduces to an exchangeable random-intercept model.

\subsubsection{Discrete-time decay structure}
The nested exchangeable structure assumes that the between-period ICC is constant for all pairs of distinct periods. In some CRXO trials, however, correlations may decrease as the temporal separation between periods increases. To represent this setting, we also consider a discrete-time decay structure in which the cluster-period random effects follow a first-order autoregressive correlation pattern. Let \(\rho_w\in(0,1)\) denote the within-period ICC and let \(r\in[0,1]\) denote the per-period decay parameter. We specify $\sigma_\gamma^2 = \frac{\rho_w}{1-\rho_w}\sigma_e^2$, so that the within-period ICC is \(\rho_w\). For two observations from the same cluster but different periods \(j\) and \(l\), the between-period ICC is
\[
\rho_{j,l}
=
\frac{\sigma_\gamma^2}{\sigma_\gamma^2+\sigma_e^2}r^{|j-l|}
=
\rho_w r^{|j-l|}.
\]
Thus, correlations between outcomes from the same cluster decay geometrically with the lag \(|j-l|\). In particular, \(\mathrm{ICC}_{j,j+1}=\rho_w r\), so \(r\) can be interpreted as the lag-one cluster autocorrelation. For periods separated by \(|j-l|\) lags, the corresponding correlation ratio is \(r^{|j-l|}\). This plays a role analogous to the CAC parameter \(\kappa\) in the nested exchangeable structure, although the two parameters imply different correlation patterns. When \(r=1\), \(\rho_{j,l}=\rho_w\) for all pairs of periods, the discrete-time decay structure reduces to an exchangeable correlation structure, analogous to the special case \(\kappa=1\) under the nested exchangeable structure.

\subsection{Random–Random Model (M1)}
\label{sec:model}
To account for the complex correlation structure inherent in longitudinal cluster designs, a "Hooper-Girling" mixed-effects model \citep{girling2016efficiency_swd, hooper2016ss_swd} can be specified for CRXO trials. This model is also widely considered in stepped wedge cluster randomized trials. The variance-component representation of this model is given in Equation~\ref{eq:M1}.

This specification separates the within-period intracluster correlation from the between-period intracluster correlation. We use the cluster autocorrelation to describe the degree to which correlations persist across different periods within the same cluster; under this model, it depends on the relative contribution of the cluster-level variance component compared with the total cluster-level and cluster-period-level variance. The random-intercept-only model considered below as M3 is therefore a reduced version of M1 obtained by setting $\sigma_\gamma^2 = 0$.

This model flexibly accounts for cluster-level correlation and additional cluster-period heterogeneity in CRXO designs. The fixed period effect \(\phi_j\) adjusts for secular period effects, while the random cluster-period component allows outcome correlations to differ within the same period versus across different periods. Such flexibility is advantageous over specifications that assume a constant correlation over time, which can lead to misspecification. As Morgan et al. \citep{morgan2017binary_crxo} showed, when there is non-negligible within-cluster correlation, failing to model this component can lead to an inflation of type I error when the outcome is binary. As a result, explicitly accounting for the cluster-period-level correlation is often necessary for valid inference.

In practice, this model is used for the design and analysis of CRXO trials, as illustrated by the following trial. The Selective Decontamination of the Digestive Tract in the Intensive Care Unit (SuDDICU) trial,\citep{suddicu2022sdd} a multicenter cluster-randomized crossover trial that investigated whether selective decontamination of the digestive tract (SDD) reduces mortality among critically ill adults across multiple ICUs, explicitly utilized this approach. In this trial, participating ICUs were randomly assigned to alternating 12-month periods of either the SDD intervention or standard care. Because the intervention was implemented ICU-wide, patient outcomes were subject to correlations both within the specific ICU and within that specific 12-month time frame. Consequently, an individual-level hierarchical logistic regression model that incorporated both a random cluster effect and a random cluster-period effect was employed.

\subsection{Fixed–Random Model (M2)}
To mitigate potential singularity issues induced by the standard random-effects structure, the fixed-random model modifies the previous specification by replacing the random cluster intercept with fixed cluster effects while retaining the random cluster-period effect. The model is specified as:
\begin{equation}
  Y_{ijk}
  = \beta_0 + \theta X_{ij}
    + \phi_j
    + \sum_{c=1}^{I} \alpha_c I_{[i=c]}
    + \gamma_{ij} + \varepsilon_{ijk},
  \label{eq:M2}
\end{equation}
where $\alpha_c$ is the fixed effect for the $c$th cluster and $I_{[i=c]}$ is the dummy variable for clusters. This specification targets robust within-cluster identification of the intervention effect by absorbing all time-invariant cluster-level heterogeneity. Such a fixed-effects perspective is particularly attractive when the number of clusters is small and residual confounding or imperfect balance across clusters is a concern. In these scenarios, standard mixed-effects assumptions, such as the independence between random effects and covariates, often become fragile.\citep{lee2022unexposed_swd, lee2024fe_swd}

This model has been adopted in cluster-randomized crossover trials when investigators seek to control for time-invariant heterogeneity across clusters while preserving the within-cluster-period dependence structure. For example, in the DOSE VF trial (Double Sequential External Defibrillation for Refractory Ventricular Fibrillation),\citep{Drennan2020DOSEVF} a cluster randomized trial with repeated crossover conducted across paramedic services, the prespecified primary analysis used a generalized linear mixed model with fixed effects for cluster and period, together with a random effect for cluster-period. By treating clusters as fixed, the analysis absorbs persistent between-cluster differences, while the random cluster-period term accounts for residual correlation among observations arising within the same cluster and period. This empirical example closely aligns with the theoretical rationale of this model and illustrates its practical value in certain scenarios.

\subsection{Random Effects Model (M3)}
The Hussey and Hughes model\citep{hussey2007swd}, which was originally developed for estimating the treatment effect in stepped-wedge CRTs, can also be employed for CRXO design. In direct contrast to models accounting for complex temporal dynamics, it assumes a simplified exchangeable correlation structure, which can be specified as,
\begin{equation}
  Y_{ijk}
  = \beta_0
    + \theta X_{ij}
    + \phi_j
    + u_i
    + \varepsilon_{ijk},
  \label{eq:M3}
\end{equation}
Compared to the random-random model (M1), this model imposes the constraint that the cluster-period variance $\sigma_\gamma^2$ is zero. As noted above, this makes M3 a reduced version of M1. By dropping the cluster-period effect, the correlation within the same cluster is forced to be constant regardless of the time separation between measurements, so that the between-period intracluster correlation equals the within-period intracluster correlation and the cluster autocorrelation is 1. The assumption of constant correlation poses significant risks of misspecification\citep{li2021mixed_swd}, and both simulation and empirical results show that failing to differentiate between the within- and between-period correlations could lead to under-coverage of confidence intervals and inflated Type I error rates.\citep{ouyang2024misspecified_swd, Kasza2019} Nevertheless, there could be incentives to have a simplified model when the number of clusters is small in a CRXO trial. At the minimum, comparing its performance against more complex specifications explicitly highlights the impact of temporal correlation structures on treatment effect estimation. 

A useful empirical example of this simplified correlation approach is provided by the Proton Pump Inhibitors vs Histamine-2 Receptor Blockers for Ulcer Prophylaxis Treatment in the Intensive Care Unit (PEPTIC) trial,\citep{peptic2020jama} an international CRXO conducted across 50 ICUs. Its primary analysis used generalized estimating equations with exchangeable working correlation and robust standard errors clustered by ICU. This strategy did not explicitly model a separate cluster-period random effect and therefore reflects a reduced correlation structure relative to more complex specifications. In post-hoc sensitivity analyses, the PEPTIC investigators also considered a generalized linear mixed model with random effects for both site and site-period. Using the completed trial data, the estimated within-period and between-period correlations were 0.0322 and 0.0317, respectively, corresponding to a cluster autocorrelation greater than 0.99; the estimated site-period variance component was very small. These results suggest little additional need for a separate site-period random effect, supporting a simpler exchangeable or site-only random-effects specification as a parsimonious analysis for PEPTIC.

\subsection{Fixed Effect Model (M4)}
In the setting with a small number of clusters, one may also consider using a least-squares dummy-variable (LSDV) approach with fixed period and fixed cluster effects:
\begin{equation}
    Y_{ijk} = \beta_{0} + \theta X_{ij} + \phi_j + \sum_{c=1}^{I} \alpha_c I_{[i=c]} + \varepsilon_{ijk},
\end{equation}
Compared to models with mixed-effects specifications, the defining characteristic is that statistical inference for the intervention effect is exclusively driven by within-cluster comparisons. By absorbing all time-invariant cluster-level confounding into the fixed cluster intercepts, the estimator becomes highly robust. In stepped-wedge settings, fixed-effects approaches have been advocated as attractive alternatives when the number of clusters is small because mixed-effects models may yield overly narrow confidence intervals and inflated Type I error. \citep{lee2022unexposed_swd, lee2024fe_swd} Fixed-effects specifications are inherently more robust to such issues and have often been adopted in real-world trials with a small number of clusters\citep{tong2025small_swd_review}. 

One real-world application of the fixed effects model analyses is still PEPTIC\citep{peptic2020jama}, which considered a generalized linear model with fixed effects for ICU in a sensitivity analysis of the primary outcome, illustrating the use of a fixed-effects specification as an alternative analysis strategy. Taken together, these considerations suggest that fixed-effects specifications can provide a useful alternative when the number of clusters is small or when distributional assumptions for cluster-level random effects are of concern.

\subsection{Cluster-Period Summary Analysis (M5)}
To simplify the potentially complicated correlation structure induced by clusters and periods, an alternative approach in the setting of a small CRXO is to model the cluster-period means rather than analyzing individual-level data directly. This approach performs analysis at the cluster-period level and can be viewed as a principled dimension-reduction strategy that avoids unstable estimation in the small CRXO setting. The cluster–period mean can be written as:
\[
  \bar Y_{ij} \equiv \frac{1}{n_{ij}} \sum_{k=1}^{n_{ij}} Y_{ijk}.
\]
Thus, the cluster-period model regresses the means on treatment, fixed period effects, and fixed cluster effects:
\begin{equation}
  \bar Y_{ij}
  = \beta_0 + \theta X_{ij}
    + \phi_j
    + \sum_{c=1}^{I} \alpha_c I_{[i=c]}
    + \eta_{ij},
  \label{eq:M5}
\end{equation}
The error \(\eta_{ij}\) captures the cluster-period-level noise. Under the individual-level data-generating process
\(Y_{ijk}= \beta_0+\theta X_{ij} + \phi_j + u_i + \gamma_{ij} + \varepsilon_{ijk}\),
the fixed cluster effects absorb \(u_i\), and the remaining cluster-period-level error can be written as
$\eta_{ij} = \gamma_{ij} + \bar\varepsilon_{ij}$, with $\mathrm{Var}(\eta_{ij})=\sigma_\gamma^2+\sigma_\epsilon^2 / {n_{ij}}$, where
$\bar\varepsilon_{ij} = n_{ij}^{-1}\sum_{k=1}^{n_{ij}}\varepsilon_{ijk}$. 
In the current implementation, the cluster-period summary model is fitted using unweighted ordinary least squares, and the variance estimator for the treatment effect is the standard model-based OLS variance estimator from the cluster-period-level regression.

Given the small cluster-period size focus in our study, we considered two inferential versions of the same cluster-period summary regression. The first version, denoted as M5a, uses the asymptotic normal critical value for inference. The second version, denoted as M5b, uses a \(t\)-distribution based on the residual degrees of freedom from the fitted cluster-period regression. Therefore, M5a and M5b have the same fitted mean model, point estimator \(\hat{\theta}\), model-based standard error, relative bias, and RMSE. The distinction between M5a and M5b lies only in the inferential step. Let $\nu = \mathrm{df}_{\mathrm{res}}$ denote the residual degrees of freedom from the fitted cluster-period regression. Under the complete, period-balanced simulation design in which every cluster contributes observations in every period, this is $\nu = IJ - (I+J)$, where \(IJ\) is the number of cluster-period observations and \(I+J\) is the number of regression parameters after accounting for the intercept, treatment indicator, \(J-1\) period indicators, and \(I-1\) cluster indicators. In the special two-period case, this reduces to $\nu = 2I - (I+2) = I-2$, which corresponds to the degrees of freedom used in the two-period cluster-level summary regression considered by Morgan et al. If \(\nu \leq 0\), the cluster-period regression is saturated, and the residual variance, standard error, and corresponding confidence interval cannot be estimated. Therefore, neither M5a nor M5b provides valid inference in this setting. In summary, M5b is included to isolate the effect of small-sample inference for the cluster-period summary regression. Comparisons between M5a and M5b therefore reflect the impact of using \(t\)-based rather than normal-based inference, rather than a difference in the fitted mean model.

Prior evidence in cluster randomized cross-over trials with a binary outcome\citep{morgan2017binary_crxo} suggests that unweighted cluster-level regression can remain robust across a wide range of correlation scenarios, although it may lose power when within cluster-period correlation is non-zero and may require adequate cluster-period size to compensate for the loss of power. Empirical CRXOs with a small to moderate number of clusters have also adopted cluster-period analysis.\citep{mcintyre2025lactated_ringer}

\subsection{Two-Period Crossover-Difference Estimators (M6)}

As an additional cluster-level analysis for two-period CRXO designs, we considered two crossover-difference estimators. Unlike M5, which retains all cluster-period means and regress them on treatment, period, and cluster fixed effects, the crossover-difference estimators first collapse each cluster into a treatment-control contrast. These estimators are directly defined only for the two-period setting, \(J=2\), where each cluster contributes exactly one intervention cluster-period and one control cluster-period.

Let \(\bar{Y}_{ij}\) denote the cluster-period mean, as in M5. In a two-period CRXO design, each cluster contributes one intervention cluster-period and one control cluster-period. For cluster \(i\), let \(\bar{Y}_{ij}\) with \(X_{ij}=1\) denote the intervention-period mean and \(\bar{Y}_{ij}\) with \(X_{ij}=0\) denote the control-period mean. The within-cluster treatment-control contrast is then $D_i = \bar{Y}_{iT}-\bar{Y}_{iC}$. Because \(D_i\) is formed within the cluster, time-invariant cluster-level heterogeneity is removed by construction.

To account for the direction of the treatment sequence, we define \(a_i=+1\) when the first-period treatment indicator is \(X_{i1}=0\), corresponding to a control-to-intervention sequence, and \(a_i=-1\) when \(X_{i1}=1\), corresponding to an intervention-to-control sequence. Under a two-period crossover design, the contrast can be written as $D_i = \theta + \delta a_i + e_i$, where \(\theta\) is the treatment effect and \(\delta\) captures the period-contrast component. Thus, \(\theta\) is estimated by the intercept in the linear regression of \(D_i\) on \(a_i\).

\subsubsection{M6a: Unweighted crossover-difference estimator}

The unweighted estimator fits $D_i = \theta + \delta a_i + e_i$ by ordinary least squares with one row per cluster. Equivalently, the implementation fits $\mathrm{lm}(D \sim a)$ and takes the intercept as the estimate of the treatment effect $\hat{\theta}_{\mathrm{F,unwt}} = \widehat{\mathrm{Intercept}}$.
The standard error is the model-based standard error of the intercept from this cluster-level regression.
Since the regression has \(I\) cluster-level observations and two regression parameters, the residual
degrees of freedom is $\nu = I-2$. The 95\% confidence interval is $\hat{\theta}_{\mathrm{F,unwt}}
\pm t_{0.975,\nu}\widehat{\mathrm{SE}}(\hat{\theta}_{\mathrm{F,unwt}})$.

\subsubsection{M6b: Size-weighted estimator}

The size-weighted option uses the same within-cluster contrast \(D_i=\bar{Y}_{iT}-\bar{Y}_{iC}\), but it weights clusters according to the harmonic mean of their two cluster-period sizes. Let \(n_{i1}\) and \(n_{i2}\) denote the numbers of individuals in cluster \(i\) during periods 1 and 2. The cluster weight is defined as
\[
w_i = \frac{2}{1/n_{i1}+1/n_{i2}}.
\]
This gives greater influence to clusters with larger effective cluster-period size across the two periods.

Let \(\mathcal{A}\) denote clusters with \(a_i=+1\), and let \(\mathcal{B}\) denote clusters with \(a_i=-1\). The weighted mean contrast in each sequence group is
\[
\bar{D}_{\mathcal{A},w}
=
\frac{\sum_{i\in \mathcal{A}} w_iD_i}{\sum_{i\in \mathcal{A}} w_i},
\quad
\bar{D}_{\mathcal{B},w}
=
\frac{\sum_{i\in \mathcal{B}} w_iD_i}{\sum_{i\in \mathcal{B}} w_i}.
\]
The size-weighted treatment-effect estimator is then
\[
\hat{\theta}_{\mathrm{F,sizewt}}
=
\frac{1}{2}
\left(
\bar{D}_{\mathcal{A},w}
+
\bar{D}_{\mathcal{B},w}
\right).
\]
In the current implementation, the variance of each weighted sequence-specific mean is estimated by
\[
\widehat{V}_{g}
=
\frac{\sum_{i\in g}w_i^2}{\left(\sum_{i\in g}w_i\right)^2}
\widehat{S}_{g,w}^{2},
\quad g\in\{\mathcal{A},\mathcal{B}\},
\]
where
\[
\widehat{S}_{g,w}^{2}
=
\frac{n_g}{n_g-1}
\frac{\sum_{i\in g}w_i(D_i-\bar{D}_{g,w})^2}{\sum_{i\in g}w_i},
\]
and \(n_g\) is the number of clusters in sequence group \(g\). The standard error is then
\[
\widehat{\mathrm{SE}}(\hat{\theta}_{\mathrm{F,sizewt}})
=
\sqrt{
\frac{1}{4}
\left(
\widehat{V}_{\mathcal{A}}
+
\widehat{V}_{\mathcal{B}}
\right)
}.
\]
The residual degrees of freedom are taken as $\nu = I-2$, matching the two-sequence, two-period contrast structure.

Both M6a and M6b require at least two valid clusters in each sequence to estimate the sequence-specific variability. They are therefore not directly applicable to designs with \(J>2\) in the current implementation. For multiple-period crossover designs, a direct extension would require defining multiple within-cluster treatment-control contrasts per cluster and accounting for their within-cluster dependence, such as using a GEE-style or cluster-robust variance approach.

\subsection{Model summary}\label{sec:sec26}

The methods described above represent practical analysis strategies for CRXO trials, ranging from flexible individual-level mixed-effects specifications to simpler fixed-effects, cluster-period, and two-period contrast-based approaches. Models M1-M5 constitute the core set of candidate methods for general CRXO settings. Specifically, M1 and M2 include a cluster-period random effect and therefore explicitly account for cluster-period heterogeneity, whereas M3 and M4 impose simpler exchangeable working structures by omitting this component. Model M5 provides a cluster-period summary analysis based on aggregated cluster-period means. For two-period CRXO designs, we additionally consider the two-period crossover-difference estimators M6a and M6b, which are specifically defined for settings in which each cluster contributes one intervention period and one control period. Because these estimators rely on a single within-cluster treatment-control contrast, they are not directly applicable to multi-period designs in the current implementation.

Although these approaches are all motivated by existing practice or methodological work, their operating characteristics in CRXO trials with very few clusters remain incompletely understood. In particular, it is unclear how robust these methods are to correlation-structure misspecification, whether explicitly modeling cluster-period heterogeneity improves inference, and whether simpler approaches may sometimes provide more reliable finite-sample performance than more highly parameterized mixed-effects models. This issue is especially relevant in CRXO settings, where limited numbers of clusters can make variance-component estimation unstable and magnify small-sample inferential problems. Moreover, correlation structures exhibiting between-period decay are frequently encountered in empirical applications,\citep{ouyang2024misspecified_swd} yet the robustness of candidate analysis methods to such temporal decay remains uncertain when the number of clusters is very small.

For this reason, we evaluate these candidate methods in a simulation study. For multi-period scenarios with $(J>2)$, we compare the core methods M1-M5 under equal cluster-period sizes. For two-period scenarios, where unequal cluster-period sizes are considered, we additionally include the two-period crossover-difference estimators M6a and M6b. Our goal is not only to assess treatment-effect estimation under different working models, but also to evaluate the extent to which model choice affects inferential performance when the underlying within-cluster correlation departs from the assumed structure.

\section{Simulation Study} 
\label{sec:simulation}
In this section, we conducted a simulation study following the ADEMP framework\citep{Morris2019} to evaluate the finite-sample performance of candidate analysis methods for CRXO trials with very few clusters. The simulation was designed to assess both statistical operating characteristics and practical feasibility under different design and correlation settings.

\paragraph{Aims.}
The simulation study has three aims: First, to evaluate the finite-sample performance of commonly used individual-level and cluster-period-level methods for estimating the treatment effect in CRXO trials with very few clusters. Second, to evaluate whether inferential performance is driven primarily by the choice between fixed and random cluster effects or by the ability of the analysis model to account for cluster-period heterogeneity. Third, to assess the robustness of each method to correlation-structure misspecification by comparing performance under nested exchangeable and discrete-time decay data-generating mechanisms.

\paragraph{Data-generating mechanisms.}
We considered CRXO designs with cross-sectional sampling. Clusters were randomized to one of two complementary treatment sequences, with equal numbers of clusters randomly allocated to each sequence, commencing in either the control or intervention condition and switching between conditions at each period. A new independent sample of individuals was simulated in each cluster in each period. Thus, each cluster contributed observations under both treatment conditions over time, allowing clusters to act as their own controls. For example, the sequences were \(ABABAB\) and \(BABABA\) for a six-period design, and \(ABA\) and \(BAB\) for a three-period design, where \(A\) and \(B\) denote the two treatment conditions. In this way, half of the clusters received each treatment condition in every period, and, over the full trial, the numbers of periods that each cluster spent under the two conditions differed by at most one. Within each cluster-period cell, for the equal-size scenarios, we assumed a common cluster-period size \(n_{ij}=50\) for all clusters and periods, with no loss to follow-up or missing data. For the unequal-size two-period scenarios, cluster-period sizes were generated independently across cluster-period cells from a zero-truncated negative binomial distribution with a target mean of 50 and a coefficient of variation \(cv_m=1\), representing a strongly imbalanced design. This specification allowed cluster-period sizes to vary both between clusters and within clusters over periods. The generated cluster-period sizes were independent of treatment assignment, period, and treatment sequence. The unequal-size two-period scenarios were also evaluated for \(I\in\{2,4,6\}\), \(\rho_w\in\{0.01,0.05,0.10\}\), and \(\kappa\in\{0.5,0.7,1.0\}\), and were used to assess the additional impact of cluster-period size imbalance on cluster-period and contrast-based analyses. For two-period crossover-difference estimators requiring sequence-specific variability estimates, scenarios with only one cluster per sequence were treated as not estimable and were reflected in the usage rate.

Under the nested exchangeable structure, outcomes are generated from Equation~\ref{eq:M1}. In addition, we also consider the discrete-time decay model proposed by Kasza \emph{et al.}\citep{Kasza2019Nonuniform}, which extends the correlation structure by allowing the between-period intracluster correlation to decay over time. The process incorporates a first-order autoregressive structure to determine the temporal dependence of cluster-period random effects. Therefore, for individual $k$ in cluster $i$ and period $j$, outcomes are generated from:
\begin{equation}
Y_{ijk}
= \beta_{0} + \theta X_{ij}
+ \phi_j
+ \gamma_{ij} + \varepsilon_{ijk}, \label{eq:dtd}
\end{equation}
The cluster-by-period random effects for cluster $i$ are $\boldsymbol{\gamma}_i=(\gamma_{i1},\ldots,\gamma_{iT})^\top
\sim \mathcal N\!\left(\mathbf 0,\ \sigma_\gamma^2\,\mathbf R_T(r)\right)$, with $\mathbf R_T(r)$ as the AR(1) correlation matrix $\{\mathbf R_T(r)\}_{jk}=r^{|j-k|}$, $0\le r\le 1$. For each combination of the design parameters described below, data were generated separately under both the nested exchangeable and discrete-time decay correlation structures.

All the simulation parameters (i.e., number of clusters, number of periods, correlation parameters, and cluster–period sizes) are chosen to reflect designs that are plausible in practice for cluster randomized crossover trials with very few clusters. The assumed values are summarized in Table~\ref{tab:sim-params}. We allowed the number of clusters $I$ to take values of 2, 4, and 6, representing trials with very small numbers of clusters. The choice of six clusters was additionally motivated by the fact that it is the smallest number of randomization units for which a permutation-based \(p\)-value of 0.05 can be obtained. This range was also motivated by Arnup et al.'s systematic review, which reported a median of 9 clusters overall (IQR, 4-21; range, 2-268) and a median of 6 clusters among trials randomizing hospitals (IQR, 2-10; range, 2-46)\citep{arnup2016appropriate}. Thus, our selected values focus on the lower range of cluster numbers observed in applied CRXO studies, where small-sample inference is expected to be most challenging. The number of periods $J$ varied between 2 and 6, covering designs ranging from simple two-period crossovers to designs with several repeated crossover periods. This choice was also informed by empirical CRXO practice.\citep{arnup2016appropriate} For the multi-period scenarios with $J>2$, each cluster-period cell contained an equal number of individuals, with $n_{ij}=50$. For the two-period scenarios, unequal cluster-period sizes were considered to evaluate methods specifically designed for two-period CRXO trials, including the two-period crossover-difference estimators. Thus, the equal-size multi-period scenarios were used to isolate the effects of the number of clusters, number of periods, and correlation structure, whereas the two-period unequal-size scenarios were used to assess the additional impact of cluster-period size imbalance. For the equal-size multi-period scenarios, the total sample size ranged according to \(I\times J\times 50\). The unequal-size two-period scenarios are described separately. To explore a wide range of correlation structures, we varied the within-period intracluster correlation $\rho_w$, spanning weak to moderate clustering. The cluster autocorrelation parameter (CAC) governing between-period correlation was set between \(0.5\) and \(1.0\). For all scenarios, we fixed the individual-level error variance \(\sigma_e^2\), so that \( \rho_w \) and the CAC parameter jointly determine the marginal ICC pattern. Crossing the grids for \(I\), \(J\), \( \rho_w \), and CAC yields \(3 \times 4 \times 3 \times 3 = 108\) design scenarios for each correlation structure.

\begin{table}[htbp]
\caption{Range of trial configuration and simulation parameter values in our simulation settings.}
\label{tab:sim-params}
\centering
\begin{tabular}{ll}
\toprule
Parameter & Values \\
\midrule
Number of clusters (\(I\)) 
    & \(2,\, 4,\, 6\) \\
Number of periods (\(J\)) 
    & \(2,\, 3,\, 5,\, 6\) \\
Cluster-period index (\((i,j)\)) 
    & \(i = 2,\, 4,\, 6;\; j = 2,\, 3,\, 5,\, 6\) \\
Cluster–period size (\(n_{ij}\)) 
    & \(50\) for \(J\ge2\); unequal sizes for \(J=2\) \\
True treatment effect (\(\theta\)) 
    & \(0,\, 0.5\) \\
Within-period ICC (\(\rho_w\)) 
    & \(0.01,\, 0.05,\, 0.10\) \\
CAC parameter (\(\kappa\) or \(r\))
    & \(0.5,\, 0.7,\, 1.0\) \\
Individual error standard deviation (\(\sigma_e\)) 
    & \(1\) \\
\bottomrule
\end{tabular}\par
\vspace{0.15cm}
\noindent\parbox{\linewidth}{\centering \footnotesize ICC: intracluster correlation coefficient; CAC: cluster autocorrelation.}
\end{table}

\paragraph{Estimand.}
The estimand of interest was the marginal treatment effect ($\theta$). Under non-null scenarios, performance was evaluated with respect to the true value ($\theta=0.5$). Under null scenarios, ($\theta=0$) was used to assess empirical type I error. Because the marginal mean model was correctly specified across all scenarios, differences in model performance were expected to arise primarily from estimation and inference under different working correlation structures rather than from bias in the target estimand.

\paragraph{Methods.}
We compared the candidate analysis methods described in Section~\ref{sec:methods}. For the multi-period scenarios with $J>2$, where cluster-period sizes were equal, we compared Models M1-M5. Models M1 and M2 include a cluster-period random effect and therefore explicitly account for cluster-period heterogeneity, differing in whether the cluster effect is treated as random or fixed. Models M3 and M4 omit the cluster-period random effect and impose a simpler exchangeable working structure, again differing in whether the cluster effect is treated as random or fixed. Model M5 analyzes cluster-period means using fixed cluster and period effects. Models M1-M3 were fitted in R using the \texttt{lme4} package with restricted maximum likelihood (REML) estimation. Models M4 and M5 were fitted by ordinary least squares using the base R \texttt{lm} function, at the individual and cluster-period levels, respectively. Model-based standard errors were used, with inference based on the normal reference distribution.

For the two-period scenarios, where unequal cluster-period sizes were considered, we additionally included the two-period crossover-difference estimators M6a and M6b. These estimators are defined specifically for two-period CRXO designs because each cluster contributes one intervention cluster-period and one control cluster-period, allowing a within-cluster treatment-control contrast to be formed directly. M6a uses an unweighted contrast estimator, whereas M6b uses a size-weighted version based on the harmonic mean of the two cluster-period sizes. The two-period crossover-difference estimators were implemented from cluster-specific crossover differences, with M6a fitted using an unweighted linear regression and M6b calculated using cluster-size weights and sequence-specific variance estimates; both used a $t$ reference distribution with $I-2$ degrees of freedom. Because these estimators are defined here only for two-period CRXO designs, they were not included in the multi-period comparisons.

\paragraph{Performance measures.}
In each simulation scenario, we generated S = 5,000 simulated data sets and estimated the intervention effect $\hat{\theta}_s$ using each model described above. \( \hat{\theta}_s \) denotes the estimated effect in the \( s \)th simulation, and \( \theta \) denotes the true effect. We report the properties of the intervention effect estimator in terms of relative bias, root mean squared error (RMSE), coverage probability (CP), and empirical type I error.

We present the relative bias (Rel Bias = $[\bar{\theta}-\theta]/\theta$) when $\theta \neq 0$ and the square root of the average squared difference between the estimated effect $\hat{\theta}_s$ and the true effect over the simulated data sets for each scenario (RMSE = $\sqrt{\sum_{s=1}^{S} [\hat{\theta}_s - \theta]^2 / S}$). 
Accordingly, for coverage probability or type I error, the Monte Carlo standard error was calculated as $\sqrt{\hat p(1-\hat p)/S}$, where \(\hat p\) is the empirical proportion. With \(S=5000\) and \(\hat p\approx0.95\), this value is approximately 0.003.

Empirical power, when \(\theta\neq 0\), and empirical type I error, when \(\theta=0\), were defined as the proportions of simulated data sets in which the method-specific two-sided 5\% test rejected \(H_0:\theta=0\). Equivalently, these quantities were computed as the proportions of method-specific 95\% confidence intervals that excluded 0. For normal-based procedures, confidence intervals were constructed using the corresponding critical value; for \(t\)-based procedures, confidence intervals were constructed using the \(t_{0.975,\nu}\) critical value with the corresponding residual degrees of freedom. Coverage probability was estimated as the proportion of method-specific 95\% confidence intervals that contained the true treatment
effect \(\theta\).

However, the above method for calculating the power can be described as an “uncorrected” approach to determining the power, where inferences are made with the normal distribution and no degrees of freedom computation is considered. 
The coverage probability is defined as $\mathrm{CP} = P\!\left( \hat\theta_{\mathrm{low}} < \theta < \hat\theta_{\mathrm{upp}} \right)$, where $\hat\theta_{\mathrm{low}}$ and $\hat\theta_{\mathrm{upp}}$ denote the lower and upper limits of the 95\% confidence interval for the treatment effect.
In the simulation study, this probability is estimated by $\widehat{\mathrm{CP}} = \sum_{s=1}^{S} \mathbb{I}\!\left(\hat\theta_{\mathrm{low}}^{(s)} < \theta < \hat\theta_{\mathrm{upp}}^{(s)}\right) / S.$

We additionally track the usage rate, defined as the proportion of simulations in which a method was successfully fitted and returned an estimable treatment-effect coefficient, and the singular rate, defined as the proportion of successful mixed-model fits with one or more random-effect variance components estimated near zero. For the mixed-effects models, these metrics are useful for assessing estimation instability and overparameterization in \texttt{lmer} fits from the \texttt{lme4} package.\citep{bates2015lme4}

\section{Simulation results}
Model feasibility differed across candidate methods. Detailed usage and singular-fit rates across all simulation settings are reported in Appendix Tables~37-42. Under the nested exchangeable data-generating mechanism (Appendix Tables~37-39), reduced usage rates were concentrated in the most highly parameterized mixed-effects model, M1, particularly in the sparsest designs. With only two clusters, the usage rate for M1 ranged from 0.595 to 0.996 across simulation settings and was lowest in the two-period designs, where it was approximately 0.60-0.76. In contrast, M1 usage was at least 0.920 with four clusters and at least 0.926 with six clusters. M2 was substantially more stable, with usage rates of at least 0.952 with two clusters and approximately 1.00 with four or six clusters. The simpler candidate methods were generally estimable in nearly all simulation settings. Similar qualitative patterns were observed under the discrete-time decay data-generating mechanism (Appendix Tables~40-42). Among successfully fitted mixed models, singular fits were also common, particularly when the number of clusters was very small or when a variance component was close to the boundary. For example, under the nested exchangeable mechanism with two clusters, CAC = 0.5, and ICC = 0.01, the singular-fit rate for M1 ranged from approximately 0.74 to 0.90 across the numbers of periods considered (Appendix Table~37). These results highlight a practical trade-off, as M1 and M2 generally provided better inferential performance by explicitly accounting for cluster-period heterogeneity, but their advantages must be considered alongside potential estimation instability in extremely sparse CRXO designs.

Regarding relative bias and precision, point estimates of the treatment effect ($\theta$) were essentially unbiased across the candidate methods. In the core multi-period scenarios, relative bias for M1-M5 remained close to zero across combinations of within-period ICC, CAC, the number of clusters, and the number of periods, with no method showing systematic over- or underestimation. Similar patterns were observed in the unequal-size two-period scenarios, where the two-period crossover-difference estimators M6a and M6b were additionally considered.

Table~\ref{tab:nodecay} shows that the RMSE generally decreased as the number of clusters and periods increased, reflecting the expected efficiency gains from additional cluster-period information. Larger values of $\rho_w$ were associated with higher RMSE, whereas differences across CAC levels were comparatively modest, indicating that within-period correlation had a stronger impact on precision than temporal persistence. Differences in RMSE between models were small relative to the overall design effects, suggesting that model choice is driven primarily by inferential properties (coverage and type~I error) rather than by point-estimation efficiency. These results also provide an assessment of correlation-structure misspecification. Under the nested exchangeable data-generating process, M1 matches the variance-component structure used to generate the data, whereas M3 and M4 impose simplified working structures that omit the cluster-period component and are therefore misspecified when the cluster-period variance is nonzero. M2 also includes a cluster-period random effect, but treats the cluster effects as fixed rather than random. Under the discrete-time decay data-generating process, none of the candidate individual-level models fully reproduces the lag-dependent correlation structure. However, M1 and M2 retain a cluster-period component and therefore partially account for additional within-cluster-period heterogeneity, whereas M3 and M4 impose simpler exchangeable working structures. M5 is best viewed as a cluster-period analysis rather than a fully specified individual-level correlation model. Across scenarios, misspecification had little impact on point-estimate bias because the marginal mean model was correctly specified, but it had clearer effects on standard errors, confidence interval coverage, and type I error, especially when the number of clusters and periods was small and the CAC was low.
\begin{figure}[t] 
\centering
\includegraphics[width=\textwidth]{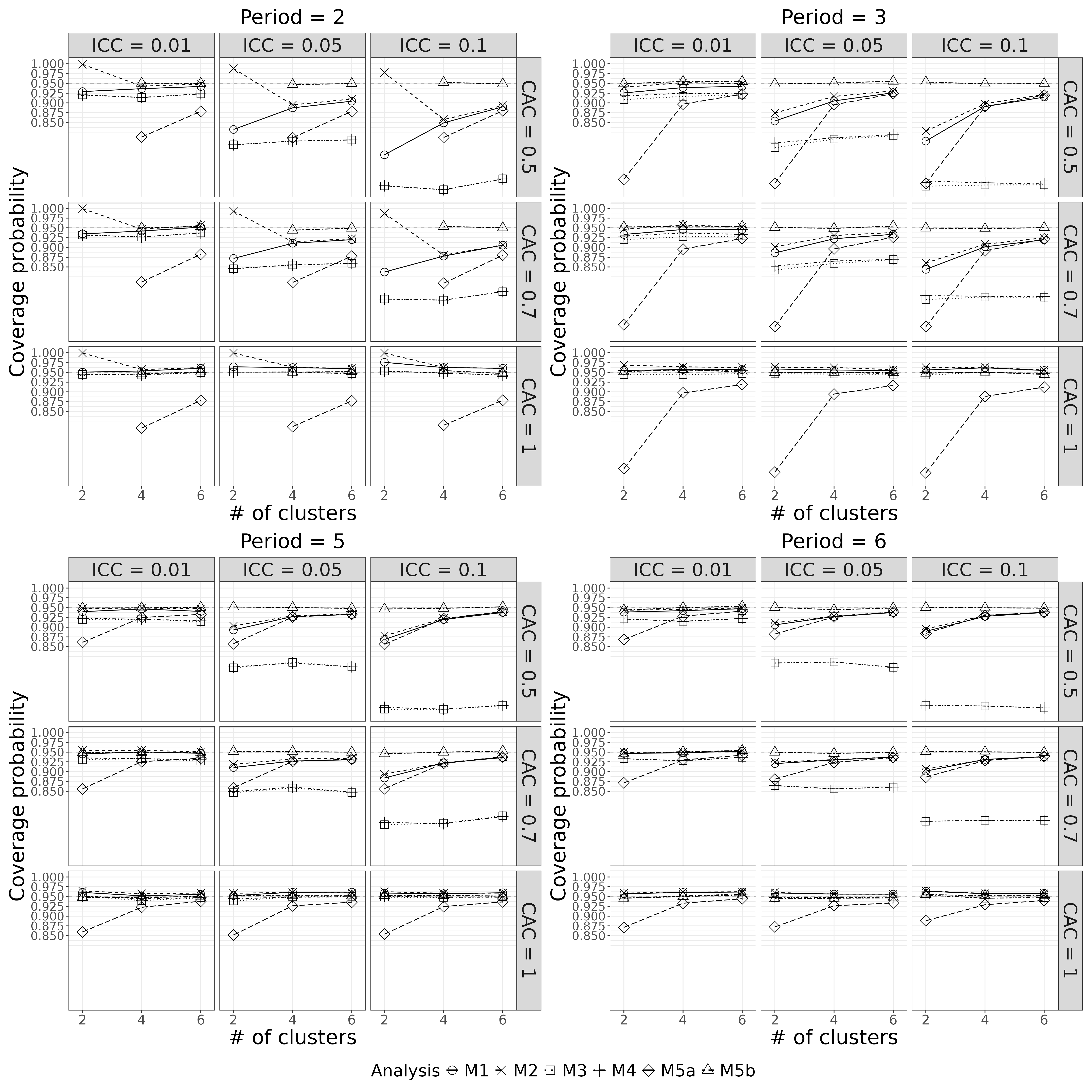}
\caption{Estimated coverage probability of nominal 95\% confidence intervals for the treatment effect (true value $\theta = 0.5$) under data generation from the nested exchangeable model.}
\label{fig:coverage}
\end{figure}

\begin{figure}[t]
\centering
\includegraphics[width=\textwidth]{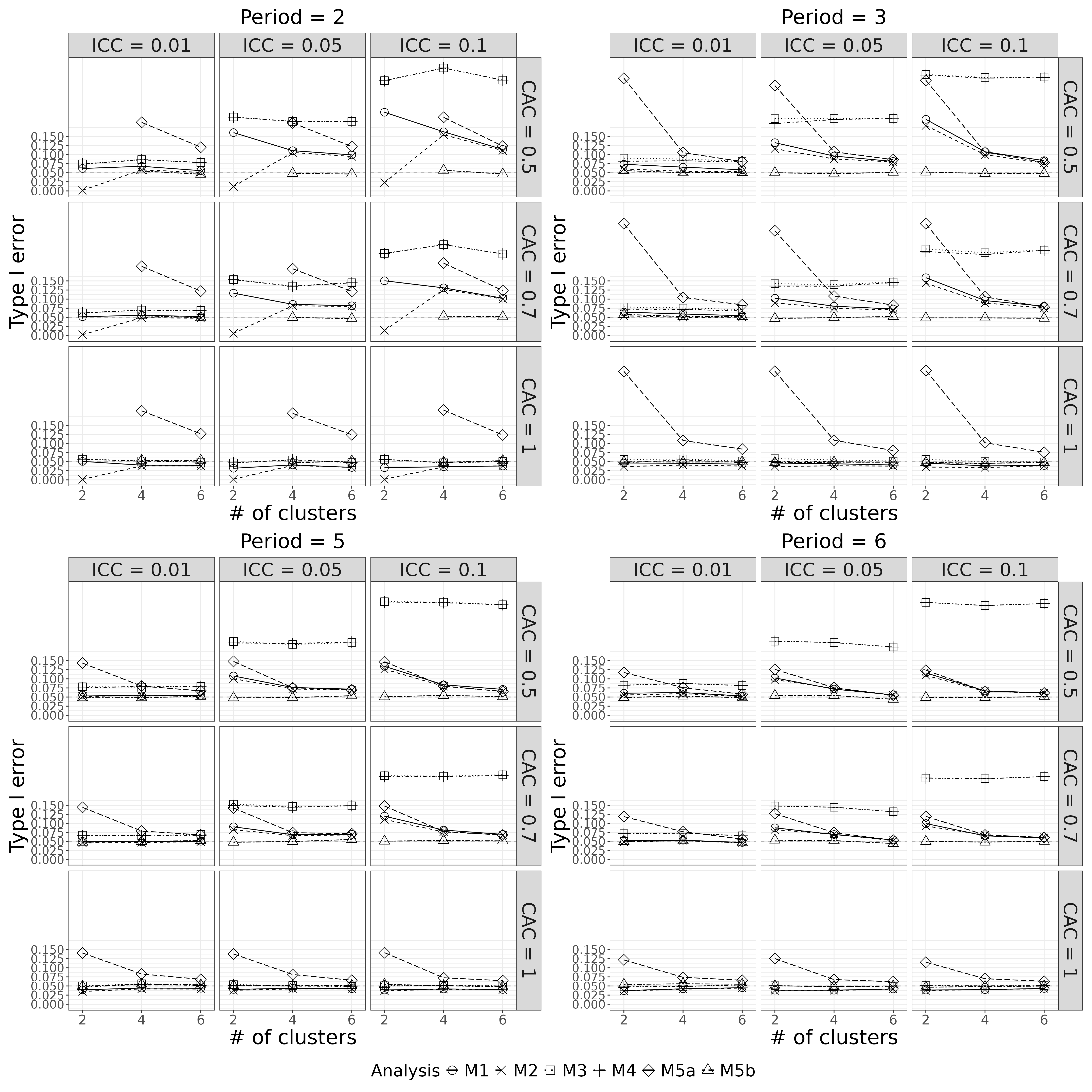}
\caption{Estimated type I error in scenarios under data generation from the nested exchangeable model.}
\label{fig:Type1Err}
\end{figure}
When the within-period intracluster correlation was low ($\rho_w=0.01$) and the cluster autocorrelation deviated from unity ($\mathrm{CAC}\neq 1$), models M1 and M2 often achieved empirical coverage rates very close to, and often slightly above, the nominal 95\% level across most of $J$ and $I$ (left-hand column of Figure~\ref{fig:coverage} for $\mathrm{CAC}=0.5$ and $0.7$). In contrast, models M3 and M4 tended to have confidence intervals with lower than nominal coverage rates in the scenarios with a smaller number of clusters and periods, such as $J=2$ and $I=2$. The coverage improved as either the number of periods or the number of clusters increased. The corresponding type~I error rates under the null hypothesis (Figure~\ref{fig:Type1Err}) mirror these patterns. Specifically, M1 and M2 remained close to the nominal 0.05 level, whereas M3 and M4 exhibited mild inflation in the smallest designs that attenuated with increasing $J$ and $I$. Model M5a performed worst when information was scarce, with noticeable under-coverage and elevated type~I error for $J=2$. However, its coverage probability increased steadily with additional periods. Surprisingly, for $J\ge 5$ and moderate numbers of clusters, M5a surpassed M3 and M4 and became comparable to M1 and M2. Consequently, the validity of normal-based inference heavily relies on substantial temporal replication (i.e., a larger number of periods) to offset the instability of the cluster-period variance estimator.

Under higher within-period correlation ($\rho_w=0.10$) combined with partial cluster persistence ($\mathrm{CAC}\neq 1$), almost all models exhibited under-coverage relative to the nominal 0.95 level except for M5b (right-hand column of Figure~\ref{fig:coverage}). For small $J$ and $I$, coverage probabilities for all models fell substantially below the nominal 0.95 level, with the largest deficiencies observed for M3 and M4, whose empirical coverage rates frequently dropped below 0.85. The associated type~I error rates in Figure~\ref{fig:Type1Err} are correspondingly inflated, particularly for M3, M4, and M5a in highly correlated settings. As the number of periods or clusters increased, coverage rates generally moved upward toward the nominal level, while type I error rates moved downward toward the nominal level. However, substantial under-coverage persisted for M3 and M4 at $\rho_w=0.10$ even in the larger scenarios. These findings indicate that the performance of M3 and M4 is particularly sensitive to strong within-period correlation when cluster effects display only partial temporal persistence.

The influence of the temporal persistence of cluster effects is most obvious by comparing the rows of Figure~\ref{fig:coverage}. When the cluster autocorrelation equaled unity ($\mathrm{CAC}=1$), the empirical coverage rates for models M1 through M4 remained consistently close to the nominal 0.95 level across all values of $\rho_w$, $J$ and $I$ (bottom row of Figure~\ref{fig:coverage}), and the corresponding type~I error rates in Figure~\ref{fig:Type1Err} were tightly centered around 0.05. This pattern suggests that these models calibrate uncertainty appropriately when cluster effects are perfectly persistent across periods. In contrast, for $\mathrm{CAC}=0.5$ or $0.7$, coverage for M3 and M4 deteriorated as $\rho_w$ increased, with the largest deficits occurring when both within-period correlation and partial persistence were present. The performance of M1 and M2 was less affected by departures from $\mathrm{CAC}=1$, with only modest reductions in coverage at high $\rho_w$. Overall, these results highlight that misspecification of the cluster-level temporal correlation is most consequential for M3 and M4, while M1 and M2 remain relatively robust.

The two versions of the cluster-period summary analysis showed distinct inferential behavior. The normal-based version M5a exhibited particularly poor performance, with low coverage and inflated Type~I error across a range of ($\rho_w$) and CAC values (Figures~\ref{fig:coverage} and~\ref{fig:Type1Err}), despite having RMSE comparable to the other methods. In the most extreme configuration, with only two clusters and two periods, coverage probability was not reported because the fitted cluster-period regression was saturated. After imposing standard identifiability constraints (e.g., setting one period and one cluster effect to zero), the total number of regression parameters is ($I+J$), leaving ($\mathrm{df}_{\mathrm{res}}=IJ-(I+J)$) residual degrees of freedom. When ($I=2$) and ($J=2$), the residual degrees of freedom reduce to zero, so the error variance, standard errors, and confidence intervals for $\hat\theta$ cannot be estimated. As $J$ increased, the coverage curves for M5a rose toward the 0.95 reference line, and its type~I error moved closer to 0.05, particularly when $I$ was at least four. In several scenarios with ($J=5, 6$) and moderate within-period correlation, M5a became comparable to M1 and M2 in terms of coverage while preserving similar RMSE. In contrast, the t-based version M5b provided much better calibration whenever sufficient residual degrees of freedom were available, with coverage often close to the nominal 95\% level. However, this improved calibration came at the cost of more conservative inference. Because the finite-sample t-critical value can be substantially larger than the normal critical value when the residual degrees of freedom are small, M5b produced wider confidence intervals and lower empirical power than the corresponding normal-based procedure. Thus, the comparison between M5a and M5b illustrates a coverage-power trade-off. Specifically, t-based inference protects against anti-conservative uncertainty quantification, whereas normal-based inference may yield higher power at the risk of under-coverage and inflated type~I error. These findings suggest that cluster-period summary analysis can be useful, but its validity depends strongly on using appropriate small-sample inference and having enough residual degrees of freedom.

\begin{table}[t]
\centering
\caption{Simulation results for data with no discrete time decay by CAC, ICC, and design parameters $(i,j)$.}
\label{tab:nodecay}

\renewcommand{\arraystretch}{1}
\setlength{\tabcolsep}{3.2pt}

\resizebox{\textwidth}{!}{%
\begin{tabular}{l
cccc cccc | cccc cccc}
\toprule

\multicolumn{1}{c}{}
& \multicolumn{8}{c}{\textbf{CAC = 0.5, ICC = 0.01}}
& \multicolumn{8}{c}{\textbf{CAC = 0.5, ICC = 0.1}} \\
\cmidrule(lr){2-9}\cmidrule(lr){10-17}

& Cov & T1E & RMSE & RelB
& Cov & T1E & RMSE & RelB
& Cov & T1E & RMSE & RelB
& Cov & T1E & RMSE & RelB \\
\cmidrule(lr){2-5}\cmidrule(lr){6-9}\cmidrule(lr){10-13}\cmidrule(lr){14-17}
\multicolumn{1}{c}{Model}
& \multicolumn{4}{c}{$(i,j)=(2,3)$} & \multicolumn{4}{c}{$(i,j)=(2,6)$}
& \multicolumn{4}{c}{$(i,j)=(2,3)$} & \multicolumn{4}{c}{$(i,j)=(2,6)$} \\

\midrule

M1 & 0.926 & 0.073 & 0.135 & -0.009 & 0.939 & 0.060 & 0.091 & 0.000 & 0.803 & 0.197 & 0.235 & -0.008 & 0.889 & 0.116 & 0.158 & 0.007 \\
M2 & 0.940 & 0.061 & 0.138 & -0.008 & 0.943 & 0.055 & 0.092 & 0.001 & 0.828 & 0.180 & 0.235 & -0.005 & 0.895 & 0.110 & 0.158 & 0.008 \\
M3 & 0.908 & 0.090 & 0.136 & -0.008 & 0.921 & 0.083 & 0.092 & 0.001 & 0.687 & 0.320 & 0.234 & -0.005 & 0.701 & 0.311 & 0.158 & 0.008 \\
M4 & 0.918 & 0.082 & 0.138 & -0.008 & 0.921 & 0.082 & 0.092 & 0.001 & 0.700 & 0.319 & 0.235 & -0.005 & 0.701 & 0.311 & 0.158 & 0.008 \\
M5a & 0.705 & 0.310 & 0.138 & -0.008 & 0.868 & 0.118 & 0.092 & 0.001 & 0.694 & 0.305 & 0.235 & -0.005 & 0.884 & 0.124 & 0.158 & 0.008 \\
M5b & 0.949 & 0.056 & 0.138 & -0.008 & 0.944 & 0.049 & 0.092 & 0.001 & 0.953 & 0.052 & 0.235 & -0.005 & 0.950 & 0.049 & 0.158 & 0.008 \\

& \multicolumn{4}{c}{$(i,j)=(6,2)$} & \multicolumn{4}{c}{$(i,j)=(6,6)$}
& \multicolumn{4}{c}{$(i,j)=(6,2)$} & \multicolumn{4}{c}{$(i,j)=(6,6)$} \\

\midrule

M1 & 0.942 & 0.056 & 0.091 &  -0.000  & 0.947 & 0.053 & 0.052 & -0.004  & 0.890 & 0.114 & 0.156 & 0.004  & 0.938 & 0.062 & 0.092 & -0.002 \\
M2 & 0.948 & 0.051 & 0.091 & -0.001  & 0.949 & 0.051 & 0.052 & -0.003  & 0.893 & 0.111 & 0.156 & 0.003  & 0.938 & 0.061 & 0.092 & -0.002 \\
M3 & 0.923 & 0.078 & 0.091 & -0.001  & 0.922 & 0.082 & 0.052 & -0.003  & 0.706 & 0.305 & 0.156 & 0.003  & 0.694 & 0.307 & 0.092 & -0.002 \\
M4 & 0.923 & 0.078 & 0.091 & -0.001  & 0.922 & 0.082 & 0.052 & -0.003  & 0.706 & 0.305 & 0.156 & 0.003  & 0.694 & 0.307 & 0.092 & -0.002 \\
M5a & 0.879 & 0.120 & 0.091 & -0.001  & 0.941 & 0.058 & 0.052 & -0.003  & 0.880 & 0.123 & 0.156 & 0.003  & 0.938 & 0.061 & 0.092 & -0.002 \\
M5b & 0.950 & 0.046 & 0.091 & -0.001 & 0.953 & 0.048 & 0.052 & -0.003 & 0.949 & 0.047 & 0.156 & 0.003 & 0.949 & 0.051 & 0.092 & -0.002 \\

\midrule

\multicolumn{1}{c}{}
& \multicolumn{8}{c}{\textbf{CAC = 1, ICC = 0.01}}
& \multicolumn{8}{c}{\textbf{CAC = 1, ICC = 0.1}} \\
\cmidrule(lr){2-9}\cmidrule(lr){10-17}

& Cov & T1E & RMSE & RelB
& Cov & T1E & RMSE & RelB
& Cov & T1E & RMSE & RelB
& Cov & T1E & RMSE & RelB \\
\cmidrule(lr){2-5}\cmidrule(lr){6-9}\cmidrule(lr){10-13}\cmidrule(lr){14-17}
\multicolumn{1}{c}{Model}
& \multicolumn{4}{c}{$(i,j)=(2,3)$} & \multicolumn{4}{c}{$(i,j)=(2,6)$}
& \multicolumn{4}{c}{$(i,j)=(2,3)$} & \multicolumn{4}{c}{$(i,j)=(2,6)$} \\

\midrule

M1 & 0.955 & 0.047 & 0.120 & 0.006 & 0.957 & 0.037 & 0.083 & 0.000 & 0.956 & 0.046 & 0.123 & 0.005 & 0.964 & 0.039 & 0.080 & -0.001 \\
M2 & 0.969 & 0.037 & 0.122 & 0.006 & 0.959 & 0.036 & 0.083 & -0.001 & 0.963 & 0.035 & 0.121 & 0.006 & 0.964 & 0.038 & 0.080 & -0.002 \\
M3 & 0.944 & 0.056 & 0.120 & 0.006 & 0.946 & 0.048 & 0.083 & -0.001 & 0.943 & 0.056 & 0.122 & 0.005 & 0.954 & 0.051 & 0.080 & -0.001 \\
M4 & 0.955 & 0.047 & 0.122 & 0.006 & 0.946 & 0.048 & 0.083 & -0.001 & 0.950 & 0.047 & 0.121 & 0.006 & 0.954 & 0.051 & 0.080 & -0.002 \\
M5a & 0.704 & 0.299 & 0.122 & 0.006 & 0.871 & 0.122 & 0.083 & -0.001 & 0.693 & 0.301 & 0.121 & 0.006 & 0.888 & 0.116 & 0.080 & -0.002 \\
M5b & 0.952 & 0.050 & 0.122 & 0.006 & 0.946 & 0.054 & 0.083 & -0.001 & 0.948 & 0.048 & 0.121 & 0.006 & 0.955 & 0.047 & 0.080 & -0.002 \\

& \multicolumn{4}{c}{$(i,j)=(6,2)$} & \multicolumn{4}{c}{$(i,j)=(6,6)$}
& \multicolumn{4}{c}{$(i,j)=(6,2)$} & \multicolumn{4}{c}{$(i,j)=(6,6)$} \\

\midrule

M1 & 0.960 & 0.040 & 0.081 & 0.001 & 0.962 & 0.045 & 0.046 & 0.001 & 0.960 & 0.038 & 0.083 & 0.001 & 0.958 & 0.043 & 0.047 & 0.001 \\
M2 & 0.962 & 0.038 & 0.081 & 0.001 & 0.962 & 0.044 & 0.046 & 0.001 & 0.960 & 0.039 & 0.083 & 0.001 & 0.958 & 0.042 & 0.047 & 0.001 \\
M3 & 0.949 & 0.049 & 0.081 & 0.001 & 0.956 & 0.053 & 0.046 & 0.001 & 0.943 & 0.051 & 0.083 & 0.001 & 0.948 & 0.051 & 0.047 & 0.001 \\
M4 & 0.949 & 0.049 & 0.081 & 0.001 & 0.956 & 0.053 & 0.046 & 0.001 & 0.943 & 0.051 & 0.083 & 0.001 & 0.948 & 0.051 & 0.047 & 0.001 \\
M5a & 0.878 & 0.127 & 0.081 & 0.001 & 0.944 & 0.066 & 0.046 & 0.001 & 0.879 & 0.124 & 0.083 & 0.001 & 0.940 & 0.063 & 0.047 & 0.001 \\
M5b & 0.951 & 0.054 & 0.081 & 0.001 & 0.954 & 0.055 & 0.046 & 0.001 & 0.947 & 0.053 & 0.083 & 0.001 & 0.952 & 0.050 & 0.047 & 0.001 \\

\bottomrule
\end{tabular}
}
\end{table}

\begin{table}[t]
\centering
\caption{Simulation results for data with discrete time decay by CAC, ICC, and design parameters $(i,j)$.}
\label{tab:decay}

\renewcommand{\arraystretch}{1}
\setlength{\tabcolsep}{3.2pt}

\resizebox{\textwidth}{!}{%
\begin{tabular}{l
cccc cccc | cccc cccc}
\toprule

\multicolumn{1}{c}{}
& \multicolumn{8}{c}{\textbf{CAC = 0.5, ICC = 0.01}}
& \multicolumn{8}{c}{\textbf{CAC = 0.5, ICC = 0.1}} \\
\cmidrule(lr){2-9}\cmidrule(lr){10-17}

& Cov & T1E & RMSE & RelB
& Cov & T1E & RMSE & RelB
& Cov & T1E & RMSE & RelB
& Cov & T1E & RMSE & RelB \\
\cmidrule(lr){2-5}\cmidrule(lr){6-9}\cmidrule(lr){10-13}\cmidrule(lr){14-17}
\multicolumn{1}{c}{Model}
& \multicolumn{4}{c}{$(i,j)=(2,3)$} & \multicolumn{4}{c}{$(i,j)=(2,6)$}
& \multicolumn{4}{c}{$(i,j)=(2,3)$} & \multicolumn{4}{c}{$(i,j)=(2,6)$} \\

\midrule

M1 & 0.934 & 0.067 & 0.133 & -0.010 & 0.945 & 0.055 & 0.090 & 0.001 &
     0.844 & 0.156 & 0.223 & -0.005 & 0.941 & 0.059 & 0.148 & 0.008\\
M2 & 0.945 & 0.057 & 0.138 & -0.009 & 0.951 & 0.050 & 0.090 & 0.001 &
     0.861 & 0.139 & 0.226 & -0.004 & 0.947 & 0.055 & 0.147 & 0.009 \\
M3 & 0.916 & 0.084 & 0.134 & -0.009 & 0.924 & 0.078 & 0.090 & 0.001 &
     0.718 & 0.292 & 0.223 & -0.004 & 0.727 & 0.273 & 0.147 & 0.009 \\
M4 & 0.922 & 0.081 & 0.138 & -0.009 & 0.924 & 0.078 & 0.090 & 0.001 &
     0.722 & 0.285 & 0.226 & -0.004 & 0.727 & 0.273 & 0.147 & 0.009 \\
M5a & 0.714 & 0.286 & 0.138 & -0.009 & 0.894 & 0.108 & 0.090 & 0.001 & 0.741 & 0.245 & 0.226 & -0.004 & 0.942 & 0.061 & 0.147 & 0.009 \\
M5b & 0.957 & 0.048 & 0.138 & -0.009 & 0.961 & 0.045 & 0.090 & 0.001 & 0.952 & 0.039 & 0.226 & -0.004 & 0.978 & 0.021 & 0.147 & 0.009 \\

& \multicolumn{4}{c}{$(i,j)=(6,2)$} & \multicolumn{4}{c}{$(i,j)=(6,6)$}
& \multicolumn{4}{c}{$(i,j)=(6,2)$} & \multicolumn{4}{c}{$(i,j)=(6,6)$} \\
\midrule
M1 & 0.945 & 0.052 & 0.092 &  0.005  & 0.958 & 0.049 & 0.052 & -0.003 & 0.894 & 0.119 & 0.160 & -0.003  & 0.978 & 0.021 & 0.086 & 0.000 \\
M2 & 0.951 & 0.048 & 0.092 &  0.005  & 0.959 & 0.045 & 0.052 & -0.003 & 0.897 & 0.114 & 0.160 & -0.003  & 0.978 & 0.021 & 0.086 & 0.000 \\
M3 & 0.921 & 0.075 & 0.092 &  0.005  & 0.927 & 0.074 & 0.052 & -0.003 & 0.690 & 0.316 & 0.160 & -0.003  & 0.731 & 0.272 & 0.086 & 0.000 \\
M4 & 0.921 & 0.075 & 0.092 &  0.005  & 0.927 & 0.074 & 0.052 & -0.003 & 0.690 & 0.316 & 0.160 & -0.003  & 0.731 & 0.272 & 0.086 & 0.000 \\
M5a & 0.880 & 0.115 & 0.092 &  0.005  & 0.952 & 0.050 & 0.052 & -0.003 & 0.886 & 0.127 & 0.160 & -0.003  & 0.978 & 0.021 & 0.086 & 0.000 \\
M5b & 0.949 & 0.048 & 0.092 & 0.005 & 0.962 & 0.040 & 0.052 & -0.003 & 0.955 & 0.056 & 0.160 & -0.003 & 0.984 & 0.016 & 0.086 & 0.000 \\
\midrule

\multicolumn{1}{c}{}
& \multicolumn{8}{c}{\textbf{CAC = 1, ICC = 0.01}}
& \multicolumn{8}{c}{\textbf{CAC = 1, ICC = 0.1}} \\
\cmidrule(lr){2-9}\cmidrule(lr){10-17}

& Cov & T1E & RMSE & RelB
& Cov & T1E & RMSE & RelB
& Cov & T1E & RMSE & RelB
& Cov & T1E & RMSE & RelB \\
\cmidrule(lr){2-5}\cmidrule(lr){6-9}\cmidrule(lr){10-13}\cmidrule(lr){14-17}
\multicolumn{1}{c}{Model}
& \multicolumn{4}{c}{$(i,j)=(2,3)$} & \multicolumn{4}{c}{$(i,j)=(2,6)$}
& \multicolumn{4}{c}{$(i,j)=(2,3)$} & \multicolumn{4}{c}{$(i,j)=(2,6)$} \\

\midrule

M1 & 0.953 & 0.046 & 0.121 & 0.008 & 0.956 & 0.041 & 0.083 &  0.001 & 0.956 & 0.043 & 0.123 & 0.001 & 0.960 & 0.043 & 0.081 &  0.000 \\
M2 & 0.964 & 0.035 & 0.122 & 0.008 & 0.958 & 0.039 & 0.083 &  0.001 & 0.962 & 0.036 & 0.121 & 0.001  & 0.962 & 0.041 & 0.081 &  0.000 \\
M3 & 0.942 & 0.056 & 0.121 & 0.008 & 0.948 & 0.051 & 0.083 &  0.001 & 0.944 & 0.055 & 0.122 & 0.001 & 0.947 & 0.054 & 0.081 &  0.000 \\
M4 & 0.949 & 0.047 & 0.123 & 0.008 & 0.948 & 0.051 & 0.083 &  0.001 & 0.950 & 0.049 & 0.121 & 0.001  & 0.947 & 0.054 & 0.081 &  0.000 \\
M5a & 0.701 & 0.302 & 0.123 & 0.008 & 0.875 & 0.124 & 0.083 & 0.001 &  0.699  &  0.304  & 0.121 & 0.001 & 0.887 & 0.120 & 0.081 &  0.000 \\
M5b & 0.947 & 0.046 & 0.123 & 0.008 & 0.948 & 0.052 & 0.083 & 0.001 & 0.951 & 0.050 & 0.121 & 0.001 & 0.955 & 0.047 & 0.081 & 0.000 \\

& \multicolumn{4}{c}{$(i,j)=(6,2)$} & \multicolumn{4}{c}{$(i,j)=(6,6)$}
& \multicolumn{4}{c}{$(i,j)=(6,2)$} & \multicolumn{4}{c}{$(i,j)=(6,6)$} \\

\midrule
M1 & 0.964 & 0.037 & 0.080 &  0.002  & 0.961 & 0.043 & 0.046 & -0.002 & 0.957 & 0.039 & 0.083 & 0.000  & 0.957 & 0.045 & 0.047 & 0.000 \\
M2 & 0.966 & 0.035 & 0.080 &  0.003  & 0.963 & 0.042 & 0.046 & -0.002 & 0.957 & 0.038 & 0.083 & -0.001  & 0.958 & 0.045 & 0.047 & 0.000 \\
M3 & 0.956 & 0.046 & 0.080 &  0.003  & 0.955 & 0.051 & 0.047 & -0.002 & 0.941 & 0.052 & 0.083 & -0.001  & 0.950 & 0.053 & 0.047 & 0.000 \\
M4 & 0.956 & 0.046 & 0.080 &  0.003  & 0.955 & 0.051 & 0.047 & -0.002 & 0.941 & 0.052 & 0.083 & -0.001  & 0.950 & 0.053 & 0.047 & 0.000 \\
M5a & 0.882 & 0.115 & 0.080 &  0.003  & 0.942 & 0.064 & 0.047 & -0.002 & 0.873 & 0.128 & 0.083 & -0.001 & 0.939 & 0.067 & 0.047 & 0.000 \\
M5b & 0.952 & 0.051 & 0.080 & 0.003 & 0.954 & 0.052 & 0.047 & -0.002 & 0.946 & 0.050 & 0.083 & -0.001 & 0.949 & 0.055 & 0.047 & 0.000 \\

\bottomrule
\end{tabular}
}
\end{table}
In addition, we examined two-period CRXO scenarios with unequal cluster-period sizes, where the two-period crossover-difference estimators M6a and M6b were also applicable. Based on the results in the appendix, overall conclusions were similar to those from the equal-size two-period scenarios. Point estimates remained approximately unbiased across methods, while inferential performance was again driven mainly by the correlation structure and the amount of information available at the cluster level. The normal-based cluster-period analysis M5a was not reliable in the two-period setting and was not estimable when the residual degrees of freedom were zero. By contrast, the t-based version M5b and the unweighted crossover-difference estimator M6a produced essentially identical results in these two-period settings, as expected from their shared within-cluster treatment-control contrast formulation. The size-weighted estimator M6b produced similar coverage and type~I error control, and in unequal-size settings, it often achieved smaller RMSE than the unweighted contrast by giving greater weight to clusters with larger effective sample sizes. However, these contrast-based t-procedures also tended to be more conservative than normal-based individual-level analyses, leading to lower empirical power in some sparse settings. These results suggest that for two-period CRXO trials with unequal cluster-period sizes, contrast-based t-inference can be a useful cluster-level alternative when at least two clusters are available in each sequence group, but this gain in coverage calibration may come at the expense of power.

Regarding the model performance under nested exchangeable and time-decay structures, holding all other scenario characteristics constant, the performance profiles under the two data-generating mechanisms were similar. Specifically, for the setting with \(\mathrm{CAC}=0.50\) and \(\mathrm{ICC}=0.01\) (Tables~\ref{tab:nodecay} and~\ref{tab:decay}; Appendix Tables), the absolute bias for all candidate models never exceeded about 0.005 under either correlation structure, and the relative bias was generally at or below about 1\% across combinations of the number of clusters and periods. Differences in RMSE between the two specifications were also very small. RMSE decreased from roughly 0.16 in the smallest design with two clusters and two periods to about 0.05 in the largest design with six clusters and six periods, and the decay and nested exchangeable structures produced RMSE values that differed by at most about 0.003-0.004 within a given design. Empirical coverage and type~I error showed somewhat greater variability across models, but there was no clear systematic shift after introducing time decay. For M1 and M2, coverage was generally close to the nominal 95\% level across most configurations, although the sparsest two-cluster, two-period setting showed some departures, including conservative behavior for M2. M3 and M4 showed modest under-coverage under both correlation structures, with coverage mostly around 0.91-0.93 and type~I error around 0.07-0.09. M5a continued to display poor performance in the smallest estimable cluster-period configuration with two clusters and three periods, with coverage around 0.71 and type~I error around 0.29-0.31 regardless of whether time decay was present. Its coverage improved steadily as the number of clusters and periods increased, reaching approximately 0.93-0.95 in larger designs, particularly when there were six clusters and five or more periods. Overall, bias, RMSE, and coverage showed only minor, non-systematic differences between the two correlation structures, which did not alter the qualitative conclusions presented above. More detailed results are presented in the Appendix Tables.

\section{Discussion}
\label{sec:discussion}
This study evaluated practical statistical modeling strategies for analyzing CRXO trials with very few clusters. A central finding is that appropriate modeling of the correlation structure, particularly the distinction between within-period and between-period correlations, is more consequential than the choice between fixed and random effects for the cluster intercept. Across a broad range of simulation scenarios, all candidate methods produced treatment effect estimates with minimal bias and similar RMSE, but their inferential performance differed meaningfully. Models that included an explicit cluster-period component generally provided the best protection against under-coverage and inflated type~I error, especially when the within-period and between-period correlations were not equal. By contrast, the simpler exchangeable models performed adequately, mainly when the true correlation structure was close to exchangeable. These findings support prioritizing models that account for cluster-period heterogeneity when analyzing CRXO trials with very few clusters. This result shifts the practical emphasis away from the choice between simple fixed-effects and random-effects. In very small CRXO trials, the more important question is whether the analysis captures period-specific cluster variation induced by the crossover design. Once this feature was modeled, the distinction between fixed and random cluster intercepts had relatively little impact on inferential performance. However, the better operating characteristics of M1 and M2 came with frequent singular fits in the smallest settings, so their statistical advantages must be interpreted alongside their computational instability. This creates a practical dilemma in very small CRXO trials, as models with more favorable inferential properties may be difficult to fit reliably, whereas computationally simpler alternatives may be more stable but can provide less reliable inference when the working correlation structure is misspecified.

The broad insights were also similar under the nested exchangeable and discrete-time decay data-generating mechanisms. Holding the other scenario characteristics constant, introducing temporal decay did not produce a clear systematic shift in bias, RMSE, coverage, or type~I error. This suggests that the principal source of inferential difficulty was not temporal decay alone, but the failure of the analysis model to distinguish within-period from between-period correlation. This finding is consistent with previous theoretical work showing that, when the true within-cluster correlation follows a discrete-time decay structure, fitting an exchangeable correlation structure can lead to confidence intervals that are too narrow.\citep{Kasza2019} Although none of the candidate individual-level models fully reproduced the lag-dependent correlation under the time-decay mechanism, the Random-Random and the Fixed-Random model retained a cluster-period component and therefore partially accommodated the additional within-cluster-period heterogeneity. Also, the results for the cluster-period summary and contrast-based analyses highlight an important trade-off between calibration and power. The normal-based cluster-period procedure was often anti-conservative when the residual degrees of freedom were small, whereas the corresponding finite-sample t-based procedure substantially improved coverage and type~I error control when enough residual degrees of freedom were available. This improved calibration came at the cost of more conservative inference and lower empirical power. Thus, cluster-period or contrast-based t-inference can be a useful alternative in sparse two-period or multi-period designs, but its performance depends on having enough cluster-level information to estimate residual or sequence-specific variability. The ROSTERS data example provided an empirical illustration of several patterns observed in the simulation study. Treatment-effect estimates varied only modestly across the candidate methods, which was almost consistent with the unbiasedness observed in the simulations. In contrast, the estimated uncertainty differed substantially. Although a single empirical dataset cannot establish whether these intervals have correct frequentist coverage, this pattern is consistent with the simulation finding that failure to account for cluster-period heterogeneity can yield greater precision in certain settings.

Several limitations of this study should be acknowledged. First, we focused on continuous outcomes analyzed using linear models. For binary, count, or time-to-event outcomes, small-sample behavior may be worse because generalized linear mixed models are more sensitive to separation, non-convergence, and bias in variance component estimation. Second, our simulations assumed balanced crossover schedules, no missing data, and no carryover effects. Most multi-period scenarios used equal cluster-period sizes, although unequal cluster-period sizes were considered in the two-period scenarios. Real CRXO trials may involve more complex imbalances, missing outcomes, unequal treatment sequences, or carryover effects. Third, our simulation scenarios considered only a limited family of correlation structures. Although the nested exchangeable and discrete-time decay models capture important features of CRXO dependence, other forms of misspecification are possible, including heteroscedasticity across periods, secular drift that differs by cluster, and treatment-effect heterogeneity. Fourth, we did not evaluate Kenward-Roger or Satterthwaite small-sample corrections. Although these approaches are commonly used to improve finite-sample inference for mixed-effects models, in preliminary implementation they introduced substantial computational burden and numerical instability in the extremely small-cluster settings considered here, particularly when variance-component estimation was near the boundary or model fits were singular. We therefore focused on the model-based inference procedures described above. Further work could investigate the performance and computational feasibility of these corrections in small CRXO trials. Alternative finite-sample procedures, including permutation-based approaches or design-based randomization inference, may warrant further study in this setting.

In conclusion, our simulations show that for CRXO trials with very few clusters, inferential validity depends primarily on whether the analysis captures the cluster-period correlation structure induced by repeated crossover, rather than solely on whether cluster intercepts are treated as fixed or random. Models that include a cluster-period random effect, such as M1 and M2, generally offer the best protection against under-coverage and inflated type~I error when cluster autocorrelation is less than one. Simpler exchangeable models, such as M3 and M4, performed adequately only when the true correlation structure was compatible with their assumptions. Cluster-period summary and contrast-based analyses can be useful when appropriate small-sample inference is used and sufficient residual degrees of freedom are available, but they may be unreliable in the most sparse designs. Overall, these results support a cautious but practical recommendation:In very small CRXO trials, analysts should prioritize modeling cluster-period heterogeneity, routinely assess convergence and singularity, and interpret results from extremely sparse designs, especially those with only two clusters, with substantial caution.

\section*{Acknowledgment}
Research in this article was supported by a Patient-Centered Outcomes Research Institute Award\textsuperscript{\textregistered} (PCORI\textsuperscript{\textregistered} Award ME-2022C2-27676). AF was supported in part by a National Health and Medical Research Council of Australia Ideas Grant ID 2037218. JK is supported by a National Health and Medical Research Council of Australia Investigator Grant ID 2033380. The statements presented in this article are solely the responsibility of the authors and do not necessarily represent the official views of PCORI\textsuperscript{\textregistered}, its Board of Governors, the Methodology Committee or NHMRC.

\section*{Supporting information}
Additional supporting information, including Web Appendices, may be found online in the supporting information tab for this article. All R code for the simulation study and software is publicly available at \url{https://github.com/qianzhesun/CRXO-Analyses}.

\printbibliography

\newpage
\appendix
\setcounter{section}{0}
\setcounter{subsection}{0}
\setcounter{subsubsection}{0}

\renewcommand{\thesection}{\arabic{section}}
\renewcommand{\thesubsection}{\thesection.\arabic{subsection}}
\renewcommand{\thesubsubsection}{\thesubsection.\arabic{subsubsection}}

\section{Additional simulation results} \label{supp:add_simu}

\subsection{Supplementary Results under the Nested Exchangeable (Non-decaying) Correlation Structure}

\begin{table}[H]
\centering
\caption{Simulation results for CAC = 0.50, ICC = 0.01 (where $i$ = \#clusters, $j$ = \#periods).}
\label{tab:cac050_icc001}

\renewcommand{\arraystretch}{1}
\setlength{\tabcolsep}{3.2pt}

\resizebox{\textwidth}{!}{%
\begin{tabular}{l
cccc cccc | cccc cccc}
\toprule

\multicolumn{1}{c}{}
& \multicolumn{8}{c}{\textbf{CAC = 0.50, ICC = 0.01}}
& \multicolumn{8}{c}{\textbf{CAC = 0.50, ICC = 0.01}} \\
\cmidrule(lr){2-9}\cmidrule(lr){10-17}

& Cov & T1E & RMSE & RelB
& Cov & T1E & RMSE & RelB
& Cov & T1E & RMSE & RelB
& Cov & T1E & RMSE & RelB \\
\cmidrule(lr){2-5}\cmidrule(lr){6-9}\cmidrule(lr){10-13}\cmidrule(lr){14-17}

\multicolumn{1}{c}{Model}
& \multicolumn{4}{c}{$(i,j)=(2,2)$} & \multicolumn{4}{c}{$(i,j)=(2,3)$}
& \multicolumn{4}{c}{$(i,j)=(2,5)$} & \multicolumn{4}{c}{$(i,j)=(2,6)$} \\
\midrule

M1 & 0.929 & 0.062 & 0.159 & 0.002 & 0.926 & 0.073 & 0.135 & -0.009 & 0.940 & 0.056 & 0.101 & 0.002 & 0.939 & 0.060 & 0.091 & 0.000 \\
M2 & 0.999 & 0.002 & 0.160 & 0.003 & 0.940 & 0.061 & 0.138 & -0.008 & 0.948 & 0.051 & 0.101 & 0.002 & 0.943 & 0.055 & 0.092 & 0.001 \\
M3 & 0.920 & 0.074 & 0.159 & 0.003 & 0.908 & 0.090 & 0.136 & -0.008 & 0.919 & 0.078 & 0.101 & 0.002 & 0.921 & 0.083 & 0.092 & 0.001 \\
M4 & 0.921 & 0.074 & 0.159 & 0.003 & 0.918 & 0.082 & 0.138 & -0.008 & 0.922 & 0.075 & 0.101 & 0.002 & 0.921 & 0.082 & 0.092 & 0.001 \\
M5a & NA & NA & 0.159 & 0.003 & 0.705 & 0.310 & 0.138 & -0.008 & 0.861 & 0.143 & 0.101 & 0.002 & 0.868 & 0.118 & 0.092 & 0.001 \\
\bottomrule
\end{tabular}
}
\end{table}

\begin{table}[H]
\centering
\caption{Simulation results for CAC = 0.50, ICC = 0.05 (where $i$ = \#clusters, $j$ = \#periods).}
\label{tab:cac050_icc005}

\renewcommand{\arraystretch}{1}
\setlength{\tabcolsep}{3.2pt}

\resizebox{\textwidth}{!}{%
\begin{tabular}{l
cccc cccc | cccc cccc}
\toprule

\multicolumn{1}{c}{}
& \multicolumn{8}{c}{\textbf{CAC = 0.50, ICC = 0.05}}
& \multicolumn{8}{c}{\textbf{CAC = 0.50, ICC = 0.05}} \\
\cmidrule(lr){2-9}\cmidrule(lr){10-17}

& Cov & T1E & RMSE & RelB
& Cov & T1E & RMSE & RelB
& Cov & T1E & RMSE & RelB
& Cov & T1E & RMSE & RelB \\
\cmidrule(lr){2-5}\cmidrule(lr){6-9}\cmidrule(lr){10-13}\cmidrule(lr){14-17}

\multicolumn{1}{c}{Model}
& \multicolumn{4}{c}{$(i,j)=(2,2)$} & \multicolumn{4}{c}{$(i,j)=(2,3)$}
& \multicolumn{4}{c}{$(i,j)=(2,5)$} & \multicolumn{4}{c}{$(i,j)=(2,6)$} \\
\midrule

M1 & 0.832 & 0.161 & 0.217 & -0.008 & 0.854 & 0.133 & 0.186 & -0.002 & 0.893 & 0.108 & 0.142 & 0.002 & 0.906 & 0.103 & 0.124 & -0.003 \\
M2 & 0.988 & 0.012 & 0.219 & -0.005 & 0.874 & 0.116 & 0.188 & 0.001 & 0.903 & 0.100 & 0.141 & 0.004 & 0.911 & 0.098 & 0.124 & -0.003 \\
M3 & 0.793 & 0.203 & 0.218 & -0.004 & 0.786 & 0.199 & 0.186 & -0.000 & 0.796 & 0.203 & 0.141 & 0.004 & 0.808 & 0.204 & 0.124 & -0.003 \\
M4 & 0.793 & 0.203 & 0.218 & -0.004 & 0.798 & 0.185 & 0.188 & 0.001 & 0.798 & 0.199 & 0.141 & 0.004 & 0.808 & 0.204 & 0.124 & -0.003 \\
M5a & NA & NA & 0.218 & -0.004 & 0.694 & 0.290 & 0.188 & 0.001 & 0.858 & 0.148 & 0.141 & 0.004 & 0.882 & 0.126 & 0.124 & -0.003 \\
\bottomrule
\end{tabular}
}
\end{table}

\begin{table}[H]
\centering
\caption{Simulation results for CAC = 0.50, ICC = 0.10 (where $i$ = \#clusters, $j$ = \#periods).}
\label{tab:cac050_icc010}

\renewcommand{\arraystretch}{1}
\setlength{\tabcolsep}{3.2pt}

\resizebox{\textwidth}{!}{%
\begin{tabular}{l
cccc cccc | cccc cccc}
\toprule

\multicolumn{1}{c}{}
& \multicolumn{8}{c}{\textbf{CAC = 0.50, ICC = 0.10}}
& \multicolumn{8}{c}{\textbf{CAC = 0.50, ICC = 0.10}} \\
\cmidrule(lr){2-9}\cmidrule(lr){10-17}

& Cov & T1E & RMSE & RelB
& Cov & T1E & RMSE & RelB
& Cov & T1E & RMSE & RelB
& Cov & T1E & RMSE & RelB \\
\cmidrule(lr){2-5}\cmidrule(lr){6-9}\cmidrule(lr){10-13}\cmidrule(lr){14-17}

\multicolumn{1}{c}{Model}
& \multicolumn{4}{c}{$(i,j)=(2,2)$} & \multicolumn{4}{c}{$(i,j)=(2,3)$}
& \multicolumn{4}{c}{$(i,j)=(2,5)$} & \multicolumn{4}{c}{$(i,j)=(2,6)$} \\
\midrule

M1 & 0.767 & 0.217 & 0.274 & -0.001 & 0.803 & 0.197 & 0.235 & -0.008 & 0.869 & 0.135 & 0.177 & 0.005 & 0.889 & 0.116 & 0.158 & 0.007 \\
M2 & 0.978 & 0.023 & 0.275 & -0.008 & 0.828 & 0.180 & 0.235 & -0.005 & 0.878 & 0.127 & 0.177 & 0.008 & 0.895 & 0.110 & 0.158 & 0.008 \\
M3 & 0.688 & 0.304 & 0.275 & -0.005 & 0.687 & 0.320 & 0.234 & -0.005 & 0.690 & 0.313 & 0.177 & 0.008 & 0.701 & 0.311 & 0.158 & 0.008 \\
M4 & 0.688 & 0.303 & 0.275 & -0.005 & 0.700 & 0.319 & 0.235 & -0.005 & 0.695 & 0.312 & 0.177 & 0.008 & 0.701 & 0.311 & 0.158 & 0.008 \\
M5a & NA & NA & 0.275 & -0.005 & 0.694 & 0.305 & 0.235 & -0.005 & 0.856 & 0.147 & 0.177 & 0.008 & 0.884 & 0.124 & 0.158 & 0.008 \\
\bottomrule
\end{tabular}
}
\end{table}

\begin{table}[H]
\centering
\caption{Simulation results for CAC = 0.70, ICC = 0.01 (where $i$ = \#clusters, $j$ = \#periods).}
\label{tab:cac070_icc001}

\renewcommand{\arraystretch}{1}
\setlength{\tabcolsep}{3.2pt}

\resizebox{\textwidth}{!}{%
\begin{tabular}{l
cccc cccc | cccc cccc}
\toprule

\multicolumn{1}{c}{}
& \multicolumn{8}{c}{\textbf{CAC = 0.70, ICC = 0.01}}
& \multicolumn{8}{c}{\textbf{CAC = 0.70, ICC = 0.01}} \\
\cmidrule(lr){2-9}\cmidrule(lr){10-17}

& Cov & T1E & RMSE & RelB
& Cov & T1E & RMSE & RelB
& Cov & T1E & RMSE & RelB
& Cov & T1E & RMSE & RelB \\
\cmidrule(lr){2-5}\cmidrule(lr){6-9}\cmidrule(lr){10-13}\cmidrule(lr){14-17}

\multicolumn{1}{c}{Model}
& \multicolumn{4}{c}{$(i,j)=(2,2)$} & \multicolumn{4}{c}{$(i,j)=(2,3)$}
& \multicolumn{4}{c}{$(i,j)=(2,5)$} & \multicolumn{4}{c}{$(i,j)=(2,6)$} \\
\midrule

M1 & 0.934 & 0.051 & 0.153 & 0.008 & 0.933 & 0.064 & 0.131 & -0.009 & 0.945 & 0.051 & 0.097 & 0.002 & 0.947 & 0.053 & 0.088 & 0.001 \\
M2 & 0.999 & 0.002 & 0.153 & 0.004 & 0.946 & 0.054 & 0.133 & -0.008 & 0.954 & 0.046 & 0.097 & 0.002 & 0.950 & 0.049 & 0.088 & 0.001 \\
M3 & 0.932 & 0.062 & 0.153 & 0.003 & 0.920 & 0.078 & 0.131 & -0.008 & 0.930 & 0.067 & 0.097 & 0.002 & 0.933 & 0.072 & 0.088 & 0.001 \\
M4 & 0.932 & 0.062 & 0.153 & 0.004 & 0.929 & 0.072 & 0.133 & -0.008 & 0.935 & 0.066 & 0.097 & 0.002 & 0.933 & 0.072 & 0.088 & 0.001 \\
M5a & NA & NA & 0.153 & 0.004 & 0.702 & 0.307 & 0.133 & -0.008 & 0.856 & 0.144 & 0.097 & 0.002 & 0.871 & 0.119 & 0.088 & 0.001 \\
\bottomrule
\end{tabular}
}
\end{table}

\begin{table}[H]
\centering
\caption{Simulation results for CAC = 0.70, ICC = 0.05 (where $i$ = \#clusters, $j$ = \#periods).}
\label{tab:cac070_icc005}

\renewcommand{\arraystretch}{1}
\setlength{\tabcolsep}{3.2pt}

\resizebox{\textwidth}{!}{%
\begin{tabular}{l
cccc cccc | cccc cccc}
\toprule

\multicolumn{1}{c}{}
& \multicolumn{8}{c}{\textbf{CAC = 0.70, ICC = 0.05}}
& \multicolumn{8}{c}{\textbf{CAC = 0.70, ICC = 0.05}} \\
\cmidrule(lr){2-9}\cmidrule(lr){10-17}

& Cov & T1E & RMSE & RelB
& Cov & T1E & RMSE & RelB
& Cov & T1E & RMSE & RelB
& Cov & T1E & RMSE & RelB \\
\cmidrule(lr){2-5}\cmidrule(lr){6-9}\cmidrule(lr){10-13}\cmidrule(lr){14-17}

\multicolumn{1}{c}{Model}
& \multicolumn{4}{c}{$(i,j)=(2,2)$} & \multicolumn{4}{c}{$(i,j)=(2,3)$}
& \multicolumn{4}{c}{$(i,j)=(2,5)$} & \multicolumn{4}{c}{$(i,j)=(2,6)$} \\
\midrule

M1 & 0.872 & 0.116 & 0.193 & -0.002 & 0.886 & 0.102 & 0.164 & -0.001 & 0.911 & 0.090 & 0.124 & 0.003 & 0.921 & 0.088 & 0.108 & -0.002 \\
M2 & 0.993 & 0.006 & 0.192 & -0.003 & 0.901 & 0.090 & 0.165 & 0.000 & 0.918 & 0.083 & 0.123 & 0.003 & 0.924 & 0.083 & 0.109 & -0.002 \\
M3 & 0.845 & 0.154 & 0.192 & -0.003 & 0.842 & 0.142 & 0.164 & -0.000 & 0.846 & 0.153 & 0.123 & 0.003 & 0.864 & 0.148 & 0.109 & -0.002 \\
M4 & 0.846 & 0.154 & 0.192 & -0.003 & 0.852 & 0.136 & 0.165 & 0.000 & 0.849 & 0.149 & 0.123 & 0.003 & 0.864 & 0.148 & 0.109 & -0.002 \\
M5a & NA & NA & 0.192 & -0.003 & 0.697 & 0.288 & 0.165 & 0.000 & 0.859 & 0.142 & 0.123 & 0.003 & 0.881 & 0.127 & 0.109 & -0.002 \\
\bottomrule
\end{tabular}
}
\end{table}

\begin{table}[H]
\centering
\caption{Simulation results for CAC = 0.70, ICC = 0.10 (where $i$ = \#clusters, $j$ = \#periods).}
\label{tab:cac070_icc010}

\renewcommand{\arraystretch}{1}
\setlength{\tabcolsep}{3.2pt}

\resizebox{\textwidth}{!}{%
\begin{tabular}{l
cccc cccc | cccc cccc}
\toprule

\multicolumn{1}{c}{}
& \multicolumn{8}{c}{\textbf{CAC = 0.70, ICC = 0.10}}
& \multicolumn{8}{c}{\textbf{CAC = 0.70, ICC = 0.10}} \\
\cmidrule(lr){2-9}\cmidrule(lr){10-17}

& Cov & T1E & RMSE & RelB
& Cov & T1E & RMSE & RelB
& Cov & T1E & RMSE & RelB
& Cov & T1E & RMSE & RelB \\
\cmidrule(lr){2-5}\cmidrule(lr){6-9}\cmidrule(lr){10-13}\cmidrule(lr){14-17}

\multicolumn{1}{c}{Model}
& \multicolumn{4}{c}{$(i,j)=(2,2)$} & \multicolumn{4}{c}{$(i,j)=(2,3)$}
& \multicolumn{4}{c}{$(i,j)=(2,5)$} & \multicolumn{4}{c}{$(i,j)=(2,6)$} \\
\midrule

M1 & 0.837 & 0.150 & 0.228 & -0.004 & 0.844 & 0.159 & 0.200 & -0.005 & 0.884 & 0.120 & 0.149 & 0.006 & 0.901 & 0.099 & 0.133 & 0.007 \\
M2 & 0.987 & 0.014 & 0.231 & -0.004 & 0.860 & 0.144 & 0.198 & -0.004 & 0.893 & 0.111 & 0.149 & 0.007 & 0.907 & 0.093 & 0.132 & 0.007 \\
M3 & 0.768 & 0.226 & 0.230 & -0.004 & 0.767 & 0.238 & 0.198 & -0.005 & 0.765 & 0.231 & 0.149 & 0.007 & 0.773 & 0.225 & 0.132 & 0.007 \\
M4 & 0.768 & 0.225 & 0.230 & -0.004 & 0.777 & 0.231 & 0.198 & -0.004 & 0.770 & 0.229 & 0.149 & 0.007 & 0.773 & 0.225 & 0.132 & 0.007 \\
M5a & NA & NA & 0.230 & -0.004 & 0.697 & 0.307 & 0.198 & -0.004 & 0.857 & 0.148 & 0.149 & 0.007 & 0.886 & 0.119 & 0.132 & 0.007 \\
\bottomrule
\end{tabular}
}
\end{table}

\begin{table}[H]
\centering
\caption{Simulation results for CAC = 1.00, ICC = 0.01 (where $i$ = \#clusters, $j$ = \#periods).}
\label{tab:cac100_icc001}

\renewcommand{\arraystretch}{1}
\setlength{\tabcolsep}{3.2pt}

\resizebox{\textwidth}{!}{%
\begin{tabular}{l
cccc cccc | cccc cccc}
\toprule

\multicolumn{1}{c}{}
& \multicolumn{8}{c}{\textbf{CAC = 1.00, ICC = 0.01}}
& \multicolumn{8}{c}{\textbf{CAC = 1.00, ICC = 0.01}} \\
\cmidrule(lr){2-9}\cmidrule(lr){10-17}

& Cov & T1E & RMSE & RelB
& Cov & T1E & RMSE & RelB
& Cov & T1E & RMSE & RelB
& Cov & T1E & RMSE & RelB \\
\cmidrule(lr){2-5}\cmidrule(lr){6-9}\cmidrule(lr){10-13}\cmidrule(lr){14-17}

\multicolumn{1}{c}{Model}
& \multicolumn{4}{c}{$(i,j)=(2,2)$} & \multicolumn{4}{c}{$(i,j)=(2,3)$}
& \multicolumn{4}{c}{$(i,j)=(2,5)$} & \multicolumn{4}{c}{$(i,j)=(2,6)$} \\
\midrule

M1 & 0.950 & 0.051 & 0.143 & 0.004 & 0.955 & 0.047 & 0.120 & 0.006 & 0.961 & 0.040 & 0.091 & 0.004 & 0.957 & 0.037 & 0.083 & 0.000 \\
M2 & 0.999 & 0.001 & 0.143 & 0.005 & 0.969 & 0.037 & 0.122 & 0.006 & 0.964 & 0.035 & 0.091 & 0.004 & 0.959 & 0.036 & 0.083 & -0.001 \\
M3 & 0.945 & 0.057 & 0.143 & 0.004 & 0.944 & 0.056 & 0.120 & 0.006 & 0.951 & 0.050 & 0.091 & 0.004 & 0.946 & 0.048 & 0.083 & -0.001 \\
M4 & 0.945 & 0.057 & 0.143 & 0.004 & 0.955 & 0.047 & 0.122 & 0.006 & 0.951 & 0.047 & 0.091 & 0.004 & 0.946 & 0.048 & 0.083 & -0.001 \\
M5a & NA & NA & 0.143 & 0.004 & 0.704 & 0.299 & 0.122 & 0.006 & 0.860 & 0.141 & 0.091 & 0.004 & 0.871 & 0.122 & 0.083 & -0.001 \\
\bottomrule
\end{tabular}
}
\end{table}

\begin{table}[H]
\centering
\caption{Simulation results for CAC = 1.00, ICC = 0.05 (where $i$ = \#clusters, $j$ = \#periods).}
\label{tab:cac100_icc005}

\renewcommand{\arraystretch}{1}
\setlength{\tabcolsep}{3.2pt}

\resizebox{\textwidth}{!}{%
\begin{tabular}{l
cccc cccc | cccc cccc}
\toprule

\multicolumn{1}{c}{}
& \multicolumn{8}{c}{\textbf{CAC = 1.00, ICC = 0.05}}
& \multicolumn{8}{c}{\textbf{CAC = 1.00, ICC = 0.05}} \\
\cmidrule(lr){2-9}\cmidrule(lr){10-17}

& Cov & T1E & RMSE & RelB
& Cov & T1E & RMSE & RelB
& Cov & T1E & RMSE & RelB
& Cov & T1E & RMSE & RelB \\
\cmidrule(lr){2-5}\cmidrule(lr){6-9}\cmidrule(lr){10-13}\cmidrule(lr){14-17}

\multicolumn{1}{c}{Model}
& \multicolumn{4}{c}{$(i,j)=(2,2)$} & \multicolumn{4}{c}{$(i,j)=(2,3)$}
& \multicolumn{4}{c}{$(i,j)=(2,5)$} & \multicolumn{4}{c}{$(i,j)=(2,6)$} \\
\midrule

M1 & 0.964 & 0.032 & 0.140 & 0.002 & 0.958 & 0.046 & 0.123 & 0.000 & 0.952 & 0.041 & 0.093 & -0.001 & 0.959 & 0.038 & 0.081 & -0.003 \\
M2 & 0.999 & 0.002 & 0.140 & -0.003 & 0.963 & 0.037 & 0.123 & 0.000 & 0.959 & 0.038 & 0.092 & -0.002 & 0.960 & 0.037 & 0.081 & -0.002 \\
M3 & 0.950 & 0.047 & 0.140 & -0.001 & 0.946 & 0.058 & 0.123 & 0.000 & 0.939 & 0.054 & 0.093 & -0.001 & 0.948 & 0.051 & 0.081 & -0.002 \\
M4 & 0.950 & 0.047 & 0.140 & -0.001 & 0.950 & 0.051 & 0.123 & 0.000 & 0.945 & 0.051 & 0.092 & -0.002 & 0.948 & 0.051 & 0.081 & -0.002 \\
M5a & NA & NA & 0.140 & -0.001 & 0.695 & 0.299 & 0.123 & 0.000 & 0.852 & 0.138 & 0.092 & -0.002 & 0.872 & 0.125 & 0.081 & -0.002 \\
\bottomrule
\end{tabular}
}
\end{table}

\begin{table}[H]
\centering
\caption{Simulation results for CAC = 1.00, ICC = 0.10 (where $i$ = \#clusters, $j$ = \#periods).}
\label{tab:cac100_icc010}

\renewcommand{\arraystretch}{1}
\setlength{\tabcolsep}{3.2pt}

\resizebox{\textwidth}{!}{%
\begin{tabular}{l
cccc cccc | cccc cccc}
\toprule

\multicolumn{1}{c}{}
& \multicolumn{8}{c}{\textbf{CAC = 1.00, ICC = 0.10}}
& \multicolumn{8}{c}{\textbf{CAC = 1.00, ICC = 0.10}} \\
\cmidrule(lr){2-9}\cmidrule(lr){10-17}

& Cov & T1E & RMSE & RelB
& Cov & T1E & RMSE & RelB
& Cov & T1E & RMSE & RelB
& Cov & T1E & RMSE & RelB \\
\cmidrule(lr){2-5}\cmidrule(lr){6-9}\cmidrule(lr){10-13}\cmidrule(lr){14-17}

\multicolumn{1}{c}{Model}
& \multicolumn{4}{c}{$(i,j)=(2,2)$} & \multicolumn{4}{c}{$(i,j)=(2,3)$}
& \multicolumn{4}{c}{$(i,j)=(2,5)$} & \multicolumn{4}{c}{$(i,j)=(2,6)$} \\
\midrule

M1 & 0.975 & 0.033 & 0.141 & 0.002 & 0.956 & 0.046 & 0.123 & 0.005 & 0.959 & 0.039 & 0.091 & -0.003 & 0.964 & 0.039 & 0.080 & -0.001 \\
M2 & 0.999 & 0.002 & 0.140 & -0.001 & 0.963 & 0.035 & 0.121 & 0.006 & 0.962 & 0.037 & 0.090 & -0.003 & 0.964 & 0.038 & 0.080 & -0.002 \\
M3 & 0.953 & 0.056 & 0.141 & 0.001 & 0.943 & 0.056 & 0.122 & 0.005 & 0.949 & 0.051 & 0.091 & -0.003 & 0.954 & 0.051 & 0.080 & -0.001 \\
M4 & 0.953 & 0.056 & 0.141 & 0.001 & 0.950 & 0.047 & 0.121 & 0.006 & 0.951 & 0.050 & 0.090 & -0.003 & 0.954 & 0.051 & 0.080 & -0.002 \\
M5a & NA & NA & 0.141 & 0.001 & 0.693 & 0.301 & 0.121 & 0.006 & 0.854 & 0.142 & 0.090 & -0.003 & 0.888 & 0.116 & 0.080 & -0.002 \\
\bottomrule
\end{tabular}
}
\end{table}

\subsection{Supplementary Results under the Discrete Time Decay Correlation Structure}

\begin{table}[H]
\centering
\caption{Simulation results for CAC = 0.50, ICC = 0.01 (where $i$ = \#clusters, $j$ = \#periods).}
\label{tab:decay_cac050_icc001}

\renewcommand{\arraystretch}{1}
\setlength{\tabcolsep}{3.2pt}

\resizebox{\textwidth}{!}{%
\begin{tabular}{l
cccc cccc | cccc cccc}
\toprule

\multicolumn{1}{c}{}
& \multicolumn{8}{c}{\textbf{CAC = 0.50, ICC = 0.01}}
& \multicolumn{8}{c}{\textbf{CAC = 0.50, ICC = 0.01}} \\
\cmidrule(lr){2-9}\cmidrule(lr){10-17}

& Cov & T1E & RMSE & RelB
& Cov & T1E & RMSE & RelB
& Cov & T1E & RMSE & RelB
& Cov & T1E & RMSE & RelB \\
\cmidrule(lr){2-5}\cmidrule(lr){6-9}\cmidrule(lr){10-13}\cmidrule(lr){14-17}

\multicolumn{1}{c}{Model}
& \multicolumn{4}{c}{$(i,j)=(2,2)$} & \multicolumn{4}{c}{$(i,j)=(2,3)$}
& \multicolumn{4}{c}{$(i,j)=(2,5)$} & \multicolumn{4}{c}{$(i,j)=(2,6)$} \\
\midrule

M1 & 0.922 & 0.077 & 0.160 & 0.008 & 0.934 & 0.067 & 0.133 & -0.010 & 0.952 & 0.051 & 0.097 & 0.001 & 0.945 & 0.055 & 0.090 & 0.001 \\
M2 & 0.998 & 0.004 & 0.158 & 0.008 & 0.945 & 0.057 & 0.138 & -0.009 & 0.958 & 0.045 & 0.099 & 0.001 & 0.951 & 0.050 & 0.090 & 0.001 \\
M3 & 0.923 & 0.082 & 0.159 & 0.008 & 0.916 & 0.084 & 0.134 & -0.009 & 0.928 & 0.077 & 0.098 & 0.001 & 0.924 & 0.078 & 0.090 & 0.001 \\
M4 & 0.924 & 0.081 & 0.159 & 0.008 & 0.922 & 0.081 & 0.138 & -0.009 & 0.931 & 0.075 & 0.099 & 0.001 & 0.924 & 0.078 & 0.090 & 0.001 \\
M5a & NA & NA & NA & NA & 0.714 & 0.286 & 0.138 & -0.009 & 0.880 & 0.120 & 0.099 & 0.001 & 0.894 & 0.108 & 0.090 & 0.001 \\
\bottomrule
\end{tabular}
}
\end{table}

\begin{table}[H]
\centering
\caption{Simulation results for CAC = 0.50, ICC = 0.05 (where $i$ = \#clusters, $j$ = \#periods).}
\label{tab:decay_cac050_icc005}

\renewcommand{\arraystretch}{1}
\setlength{\tabcolsep}{3.2pt}

\resizebox{\textwidth}{!}{%
\begin{tabular}{l
cccc cccc | cccc cccc}
\toprule

\multicolumn{1}{c}{}
& \multicolumn{8}{c}{\textbf{CAC = 0.50, ICC = 0.05}}
& \multicolumn{8}{c}{\textbf{CAC = 0.50, ICC = 0.05}} \\
\cmidrule(lr){2-9}\cmidrule(lr){10-17}

& Cov & T1E & RMSE & RelB
& Cov & T1E & RMSE & RelB
& Cov & T1E & RMSE & RelB
& Cov & T1E & RMSE & RelB \\
\cmidrule(lr){2-5}\cmidrule(lr){6-9}\cmidrule(lr){10-13}\cmidrule(lr){14-17}

\multicolumn{1}{c}{Model}
& \multicolumn{4}{c}{$(i,j)=(2,2)$} & \multicolumn{4}{c}{$(i,j)=(2,3)$}
& \multicolumn{4}{c}{$(i,j)=(2,5)$} & \multicolumn{4}{c}{$(i,j)=(2,6)$} \\
\midrule

M1 & 0.840 & 0.148 & 0.216 & -0.001 & 0.889 & 0.109 & 0.174 & -0.001 & 0.927 & 0.074 & 0.130 & -0.000 & 0.940 & 0.067 & 0.118 & -0.000 \\
M2 & 0.991 & 0.012 & 0.213 & 0.004 & 0.901 & 0.096 & 0.176 & -0.000 & 0.938 & 0.067 & 0.131 & 0.003 & 0.946 & 0.061 & 0.117 & -0.001 \\
M3 & 0.803 & 0.189 & 0.213 & 0.003 & 0.818 & 0.175 & 0.174 & 0.000 & 0.826 & 0.174 & 0.131 & 0.003 & 0.836 & 0.177 & 0.117 & -0.001 \\
M4 & 0.803 & 0.189 & 0.213 & 0.003 & 0.823 & 0.170 & 0.176 & -0.000 & 0.827 & 0.174 & 0.131 & 0.003 & 0.836 & 0.177 & 0.117 & -0.001 \\
M5a & NA & NA & NA & NA & 0.736 & 0.255 & 0.176 & -0.000 & 0.901 & 0.100 & 0.131 & 0.003 & 0.926 & 0.078 & 0.117 & -0.001 \\
\bottomrule
\end{tabular}
}
\end{table}

\begin{table}[H]
\centering
\caption{Simulation results for CAC = 0.50, ICC = 0.10 (where $i$ = \#clusters, $j$ = \#periods).}
\label{tab:decay_cac050_icc010}

\renewcommand{\arraystretch}{1}
\setlength{\tabcolsep}{3.2pt}

\resizebox{\textwidth}{!}{%
\begin{tabular}{l
cccc cccc | cccc cccc}
\toprule

\multicolumn{1}{c}{}
& \multicolumn{8}{c}{\textbf{CAC = 0.50, ICC = 0.10}}
& \multicolumn{8}{c}{\textbf{CAC = 0.50, ICC = 0.10}} \\
\cmidrule(lr){2-9}\cmidrule(lr){10-17}

& Cov & T1E & RMSE & RelB
& Cov & T1E & RMSE & RelB
& Cov & T1E & RMSE & RelB
& Cov & T1E & RMSE & RelB \\
\cmidrule(lr){2-5}\cmidrule(lr){6-9}\cmidrule(lr){10-13}\cmidrule(lr){14-17}

\multicolumn{1}{c}{Model}
& \multicolumn{4}{c}{$(i,j)=(2,2)$} & \multicolumn{4}{c}{$(i,j)=(2,3)$}
& \multicolumn{4}{c}{$(i,j)=(2,5)$} & \multicolumn{4}{c}{$(i,j)=(2,6)$} \\
\midrule

M1 & 0.767 & 0.232 & 0.275 & -0.013 & 0.844 & 0.156 & 0.223 & -0.005 & 0.923 & 0.078 & 0.161 & 0.010 & 0.941 & 0.059 & 0.148 & 0.008 \\
M2 & 0.973 & 0.025 & 0.276 & 0.003 & 0.861 & 0.139 & 0.226 & -0.004 & 0.929 & 0.068 & 0.162 & 0.012 & 0.947 & 0.055 & 0.147 & 0.009 \\
M3 & 0.688 & 0.309 & 0.276 & 0.004 & 0.718 & 0.292 & 0.223 & -0.004 & 0.729 & 0.270 & 0.162 & 0.011 & 0.727 & 0.273 & 0.147 & 0.009 \\
M4 & 0.689 & 0.309 & 0.276 & 0.004 & 0.722 & 0.285 & 0.226 & -0.004 & 0.732 & 0.270 & 0.162 & 0.012 & 0.727 & 0.273 & 0.147 & 0.009 \\
M5a & NA & NA & NA & NA & 0.741 & 0.245 & 0.226 & -0.004 & 0.915 & 0.083 & 0.162 & 0.012 & 0.942 & 0.061 & 0.147 & 0.009 \\
\bottomrule
\end{tabular}
}
\end{table}

\begin{table}[H]
\centering
\caption{Simulation results for CAC = 0.70, ICC = 0.01 (where $i$ = \#clusters, $j$ = \#periods).}
\label{tab:decay_cac070_icc001}

\renewcommand{\arraystretch}{1}
\setlength{\tabcolsep}{3.2pt}

\resizebox{\textwidth}{!}{%
\begin{tabular}{l
cccc cccc | cccc cccc}
\toprule

\multicolumn{1}{c}{}
& \multicolumn{8}{c}{\textbf{CAC = 0.70, ICC = 0.01}}
& \multicolumn{8}{c}{\textbf{CAC = 0.70, ICC = 0.01}} \\
\cmidrule(lr){2-9}\cmidrule(lr){10-17}

& Cov & T1E & RMSE & RelB
& Cov & T1E & RMSE & RelB
& Cov & T1E & RMSE & RelB
& Cov & T1E & RMSE & RelB \\
\cmidrule(lr){2-5}\cmidrule(lr){6-9}\cmidrule(lr){10-13}\cmidrule(lr){14-17}

\multicolumn{1}{c}{Model}
& \multicolumn{4}{c}{$(i,j)=(2,2)$} & \multicolumn{4}{c}{$(i,j)=(2,3)$}
& \multicolumn{4}{c}{$(i,j)=(2,5)$} & \multicolumn{4}{c}{$(i,j)=(2,6)$} \\
\midrule

M1 & 0.936 & 0.061 & 0.154 & 0.007 & 0.941 & 0.061 & 0.129 & -0.010 & 0.957 & 0.046 & 0.094 & 0.001 & 0.952 & 0.048 & 0.086 & 0.002 \\
M2 & 0.998 & 0.002 & 0.152 & 0.007 & 0.950 & 0.050 & 0.132 & -0.008 & 0.962 & 0.040 & 0.095 & 0.002 & 0.956 & 0.044 & 0.086 & 0.001 \\
M3 & 0.935 & 0.067 & 0.152 & 0.008 & 0.926 & 0.077 & 0.129 & -0.008 & 0.937 & 0.066 & 0.094 & 0.001 & 0.936 & 0.068 & 0.086 & 0.001 \\
M4 & 0.935 & 0.067 & 0.152 & 0.008 & 0.932 & 0.069 & 0.132 & -0.008 & 0.941 & 0.063 & 0.095 & 0.002 & 0.936 & 0.068 & 0.086 & 0.001 \\
M5a & NA & NA & NA & NA & 0.710 & 0.285 & 0.132 & -0.008 & 0.878 & 0.120 & 0.095 & 0.002 & 0.897 & 0.106 & 0.086 & 0.001 \\
\bottomrule
\end{tabular}
}
\end{table}

\begin{table}[H]
\centering
\caption{Simulation results for CAC = 0.70, ICC = 0.05 (where $i$ = \#clusters, $j$ = \#periods).}
\label{tab:decay_cac070_icc005}

\renewcommand{\arraystretch}{1}
\setlength{\tabcolsep}{3.2pt}

\resizebox{\textwidth}{!}{%
\begin{tabular}{l
cccc cccc | cccc cccc}
\toprule

\multicolumn{1}{c}{}
& \multicolumn{8}{c}{\textbf{CAC = 0.70, ICC = 0.05}}
& \multicolumn{8}{c}{\textbf{CAC = 0.70, ICC = 0.05}} \\
\cmidrule(lr){2-9}\cmidrule(lr){10-17}

& Cov & T1E & RMSE & RelB
& Cov & T1E & RMSE & RelB
& Cov & T1E & RMSE & RelB
& Cov & T1E & RMSE & RelB \\
\cmidrule(lr){2-5}\cmidrule(lr){6-9}\cmidrule(lr){10-13}\cmidrule(lr){14-17}

\multicolumn{1}{c}{Model}
& \multicolumn{4}{c}{$(i,j)=(2,2)$} & \multicolumn{4}{c}{$(i,j)=(2,3)$}
& \multicolumn{4}{c}{$(i,j)=(2,5)$} & \multicolumn{4}{c}{$(i,j)=(2,6)$} \\
\midrule

M1 & 0.884 & 0.106 & 0.188 & -0.002 & 0.916 & 0.080 & 0.155 & -0.001 & 0.948 & 0.056 & 0.115 & 0.003 & 0.948 & 0.055 & 0.105 & -0.001 \\
M2 & 0.994 & 0.007 & 0.188 & 0.003 & 0.927 & 0.068 & 0.155 & 0.001 & 0.953 & 0.053 & 0.115 & 0.004 & 0.953 & 0.051 & 0.105 & -0.001 \\
M3 & 0.856 & 0.137 & 0.188 & 0.003 & 0.871 & 0.126 & 0.153 & 0.000 & 0.877 & 0.123 & 0.115 & 0.004 & 0.873 & 0.132 & 0.105 & -0.001 \\
M4 & 0.857 & 0.137 & 0.188 & 0.003 & 0.876 & 0.117 & 0.155 & 0.001 & 0.878 & 0.123 & 0.115 & 0.004 & 0.873 & 0.132 & 0.105 & -0.001 \\
M5a & NA & NA & NA & NA & 0.748 & 0.262 & 0.155 & 0.001 & 0.903 & 0.094 & 0.115 & 0.004 & 0.928 & 0.077 & 0.105 & -0.001 \\
\bottomrule
\end{tabular}
}
\end{table}

\begin{table}[H]
\centering
\caption{Simulation results for CAC = 0.70, ICC = 0.10 (where $i$ = \#clusters, $j$ = \#periods).}
\label{tab:decay_cac070_icc010}

\renewcommand{\arraystretch}{1}
\setlength{\tabcolsep}{3.2pt}

\resizebox{\textwidth}{!}{%
\begin{tabular}{l
cccc cccc | cccc cccc}
\toprule

\multicolumn{1}{c}{}
& \multicolumn{8}{c}{\textbf{CAC = 0.70, ICC = 0.10}}
& \multicolumn{8}{c}{\textbf{CAC = 0.70, ICC = 0.10}} \\
\cmidrule(lr){2-9}\cmidrule(lr){10-17}

& Cov & T1E & RMSE & RelB
& Cov & T1E & RMSE & RelB
& Cov & T1E & RMSE & RelB
& Cov & T1E & RMSE & RelB \\
\cmidrule(lr){2-5}\cmidrule(lr){6-9}\cmidrule(lr){10-13}\cmidrule(lr){14-17}

\multicolumn{1}{c}{Model}
& \multicolumn{4}{c}{$(i,j)=(2,2)$} & \multicolumn{4}{c}{$(i,j)=(2,3)$}
& \multicolumn{4}{c}{$(i,j)=(2,5)$} & \multicolumn{4}{c}{$(i,j)=(2,6)$} \\
\midrule

M1 & 0.844 & 0.156 & 0.231 & 0.002 & 0.883 & 0.111 & 0.187 & -0.006 & 0.941 & 0.055 & 0.135 & 0.008 & 0.949 & 0.049 & 0.126 & 0.007 \\
M2 & 0.987 & 0.012 & 0.232 & 0.002 & 0.895 & 0.100 & 0.187 & -0.003 & 0.947 & 0.050 & 0.136 & 0.010 & 0.953 & 0.045 & 0.126 & 0.007 \\
M3 & 0.769 & 0.223 & 0.232 & 0.003 & 0.796 & 0.207 & 0.186 & -0.004 & 0.813 & 0.185 & 0.135 & 0.010 & 0.797 & 0.209 & 0.126 & 0.007 \\
M4 & 0.769 & 0.222 & 0.232 & 0.003 & 0.804 & 0.199 & 0.187 & -0.003 & 0.817 & 0.185 & 0.136 & 0.010 & 0.797 & 0.209 & 0.126 & 0.007 \\
M5a & NA & NA & NA & NA & 0.746 & 0.240 & 0.187 & -0.003 & 0.925 & 0.077 & 0.136 & 0.010 & 0.940 & 0.055 & 0.126 & 0.007 \\
\bottomrule
\end{tabular}
}
\end{table}

\begin{table}[H]
\centering
\caption{Simulation results for CAC = 1.00, ICC = 0.01 (where $i$ = \#clusters, $j$ = \#periods).}
\label{tab:decay_cac100_icc001}

\renewcommand{\arraystretch}{1}
\setlength{\tabcolsep}{3.2pt}

\resizebox{\textwidth}{!}{%
\begin{tabular}{l
cccc cccc | cccc cccc}
\toprule

\multicolumn{1}{c}{}
& \multicolumn{8}{c}{\textbf{CAC = 1.00, ICC = 0.01}}
& \multicolumn{8}{c}{\textbf{CAC = 1.00, ICC = 0.01}} \\
\cmidrule(lr){2-9}\cmidrule(lr){10-17}

& Cov & T1E & RMSE & RelB
& Cov & T1E & RMSE & RelB
& Cov & T1E & RMSE & RelB
& Cov & T1E & RMSE & RelB \\
\cmidrule(lr){2-5}\cmidrule(lr){6-9}\cmidrule(lr){10-13}\cmidrule(lr){14-17}

\multicolumn{1}{c}{Model}
& \multicolumn{4}{c}{$(i,j)=(2,2)$} & \multicolumn{4}{c}{$(i,j)=(2,3)$}
& \multicolumn{4}{c}{$(i,j)=(2,5)$} & \multicolumn{4}{c}{$(i,j)=(2,6)$} \\
\midrule

M1 & 0.952 & 0.053 & 0.143 & 0.004 & 0.953 & 0.046 & 0.121 & 0.008 & 0.959 & 0.037 & 0.091 & 0.001 & 0.956 & 0.041 & 0.083 & 0.001 \\
M2 & 0.998 & 0.002 & 0.142 & 0.002 & 0.964 & 0.035 & 0.122 & 0.008 & 0.959 & 0.031 & 0.091 & 0.001 & 0.958 & 0.039 & 0.083 & 0.001 \\
M3 & 0.947 & 0.057 & 0.143 & 0.003 & 0.942 & 0.056 & 0.121 & 0.008 & 0.947 & 0.048 & 0.091 & 0.001 & 0.948 & 0.051 & 0.083 & 0.001 \\
M4 & 0.947 & 0.057 & 0.143 & 0.003 & 0.949 & 0.047 & 0.123 & 0.008 & 0.946 & 0.044 & 0.091 & 0.001 & 0.948 & 0.051 & 0.083 & 0.001 \\
M5a & NA & NA & NA & NA & 0.701 & 0.302 & 0.123 & 0.008 & 0.853 & 0.129 & 0.091 & 0.001 & 0.875 & 0.124 & 0.083 & 0.001 \\
\bottomrule
\end{tabular}
}
\end{table}

\begin{table}[H]
\centering
\caption{Simulation results for CAC = 1.00, ICC = 0.05 (where $i$ = \#clusters, $j$ = \#periods).}
\label{tab:decay_cac100_icc005}

\renewcommand{\arraystretch}{1}
\setlength{\tabcolsep}{3.2pt}

\resizebox{\textwidth}{!}{%
\begin{tabular}{l
cccc cccc | cccc cccc}
\toprule

\multicolumn{1}{c}{}
& \multicolumn{8}{c}{\textbf{CAC = 1.00, ICC = 0.05}}
& \multicolumn{8}{c}{\textbf{CAC = 1.00, ICC = 0.05}} \\
\cmidrule(lr){2-9}\cmidrule(lr){10-17}

& Cov & T1E & RMSE & RelB
& Cov & T1E & RMSE & RelB
& Cov & T1E & RMSE & RelB
& Cov & T1E & RMSE & RelB \\
\cmidrule(lr){2-5}\cmidrule(lr){6-9}\cmidrule(lr){10-13}\cmidrule(lr){14-17}

\multicolumn{1}{c}{Model}
& \multicolumn{4}{c}{$(i,j)=(2,2)$} & \multicolumn{4}{c}{$(i,j)=(2,3)$}
& \multicolumn{4}{c}{$(i,j)=(2,5)$} & \multicolumn{4}{c}{$(i,j)=(2,6)$} \\
\midrule

M1 & 0.966 & 0.044 & 0.140 & -0.002 & 0.950 & 0.043 & 0.125 & 0.001 & 0.957 & 0.036 & 0.093 & 0.000 & 0.965 & 0.040 & 0.080 & -0.002 \\
M2 & 0.999 & 0.001 & 0.141 & -0.000 & 0.962 & 0.035 & 0.124 & 0.001 & 0.959 & 0.034 & 0.092 & 0.000 & 0.965 & 0.038 & 0.080 & -0.002 \\
M3 & 0.948 & 0.050 & 0.140 & -0.001 & 0.939 & 0.053 & 0.124 & 0.000 & 0.946 & 0.046 & 0.092 & 0.000 & 0.954 & 0.052 & 0.080 & -0.002 \\
M4 & 0.948 & 0.050 & 0.140 & -0.001 & 0.949 & 0.046 & 0.124 & 0.001 & 0.946 & 0.044 & 0.092 & 0.000 & 0.954 & 0.052 & 0.080 & -0.002 \\
M5a & NA & NA & NA & NA & 0.699 & 0.305 & 0.124 & 0.001 & 0.858 & 0.133 & 0.092 & 0.000 & 0.884 & 0.118 & 0.080 & -0.002 \\
\bottomrule
\end{tabular}
}
\end{table}

\begin{table}[H]
\centering
\caption{Simulation results for CAC = 1.00, ICC = 0.10 (where $i$ = \#clusters, $j$ = \#periods).}
\label{tab:decay_cac100_icc010}

\renewcommand{\arraystretch}{1}
\setlength{\tabcolsep}{3.2pt}

\resizebox{\textwidth}{!}{%
\begin{tabular}{l
cccc cccc | cccc cccc}
\toprule

\multicolumn{1}{c}{}
& \multicolumn{8}{c}{\textbf{CAC = 1.00, ICC = 0.10}}
& \multicolumn{8}{c}{\textbf{CAC = 1.00, ICC = 0.10}} \\
\cmidrule(lr){2-9}\cmidrule(lr){10-17}

& Cov & T1E & RMSE & RelB
& Cov & T1E & RMSE & RelB
& Cov & T1E & RMSE & RelB
& Cov & T1E & RMSE & RelB \\
\cmidrule(lr){2-5}\cmidrule(lr){6-9}\cmidrule(lr){10-13}\cmidrule(lr){14-17}

\multicolumn{1}{c}{Model}
& \multicolumn{4}{c}{$(i,j)=(2,2)$} & \multicolumn{4}{c}{$(i,j)=(2,3)$}
& \multicolumn{4}{c}{$(i,j)=(2,5)$} & \multicolumn{4}{c}{$(i,j)=(2,6)$} \\
\midrule

M1 & 0.971 & 0.035 & 0.141 & 0.003 & 0.956 & 0.043 & 0.123 & 0.001 & 0.964 & 0.042 & 0.091 & -0.005 & 0.960 & 0.043 & 0.081 & -0.000 \\
M2 & 0.999 & 0.002 & 0.141 & 0.001 & 0.962 & 0.036 & 0.121 & 0.001 & 0.968 & 0.039 & 0.091 & -0.005 & 0.962 & 0.041 & 0.081 & -0.000 \\
M3 & 0.952 & 0.059 & 0.141 & -0.000 & 0.944 & 0.055 & 0.122 & 0.001 & 0.953 & 0.054 & 0.091 & -0.005 & 0.947 & 0.054 & 0.081 & -0.000 \\
M4 & 0.952 & 0.059 & 0.141 & -0.000 & 0.950 & 0.049 & 0.121 & 0.001 & 0.955 & 0.053 & 0.091 & -0.005 & 0.947 & 0.054 & 0.081 & -0.000 \\
M5a & NA & NA & NA & NA & 0.699 & 0.304 & 0.121 & 0.001 & 0.857 & 0.146 & 0.091 & -0.005 & 0.887 & 0.120 & 0.081 & -0.000 \\
\bottomrule
\end{tabular}
}
\end{table}

\subsection{Supplementary Results under no Time Decaying Correlation Structure (unequal cluster-period size)}

\begin{table}[H]
\centering
\caption{Simulation results for CAC = 0.50, ICC = 0.01 (where $i$ = \#clusters, $j$ = \#periods).}
\label{tab:size_mode_between_and_within_cv1_cac050_icc001_period2}

\renewcommand{\arraystretch}{1.15}
\setlength{\tabcolsep}{4.5pt}

\resizebox{0.925\textwidth}{!}{%
\begin{tabular}{l cccc cccc cccc}
\toprule

\multicolumn{1}{c}{Model}
& \multicolumn{4}{c}{$(i,j)=(2,2)$}
& \multicolumn{4}{c}{$(i,j)=(4,2)$}
& \multicolumn{4}{c}{$(i,j)=(6,2)$} \\
\cmidrule(lr){2-5}\cmidrule(lr){6-9}\cmidrule(lr){10-13}

& Cov & T1E & RMSE & RelB
& Cov & T1E & RMSE & RelB
& Cov & T1E & RMSE & RelB \\
\midrule

M1 & 0.924 & 0.077 & 0.243 & 0.004 & 0.925 & 0.071 & 0.145 & -0.006 & 0.930 & 0.064 & 0.110 & 0.000 \\
M2 & 0.991 & 0.008 & 0.283 & 0.010 & 0.951 & 0.053 & 0.171 & -0.011 & 0.944 & 0.049 & 0.126 & -0.001 \\
M3 & 0.913 & 0.091 & 0.250 & 0.005 & 0.906 & 0.087 & 0.144 & -0.006 & 0.914 & 0.082 & 0.111 & 0.000 \\
M4 & 0.932 & 0.073 & 0.283 & 0.008 & 0.927 & 0.076 & 0.163 & -0.011 & 0.923 & 0.070 & 0.124 & -0.001 \\
M5a & NA & NA & 0.283 & 0.008 & 0.832 & 0.175 & 0.205 & -0.011 & 0.886 & 0.109 & 0.164 & -0.001 \\
M5b & NA & NA & NA & NA & 0.961 & 0.045 & 0.205 & -0.011 & 0.957 & 0.035 & 0.164 & -0.001 \\
M6a & NA & NA & NA & NA & 0.961 & 0.045 & 0.205 & -0.011 & 0.957 & 0.035 & 0.164 & -0.001 \\
M6b & NA & NA & NA & NA & 0.955 & 0.048 & 0.163 & -0.011 & 0.957 & 0.034 & 0.124 & -0.001 \\
\bottomrule
\end{tabular}
}
\end{table}

\begin{table}[H]
\centering
\caption{Simulation results for CAC = 0.50, ICC = 0.05 (where $i$ = \#clusters, $j$ = \#periods).}
\label{tab:size_mode_between_and_within_cv1_cac050_icc005_period2}

\renewcommand{\arraystretch}{1.15}
\setlength{\tabcolsep}{4.5pt}

\resizebox{0.925\textwidth}{!}{%
\begin{tabular}{l cccc cccc cccc}
\toprule

\multicolumn{1}{c}{Model}
& \multicolumn{4}{c}{$(i,j)=(2,2)$}
& \multicolumn{4}{c}{$(i,j)=(4,2)$}
& \multicolumn{4}{c}{$(i,j)=(6,2)$} \\
\cmidrule(lr){2-5}\cmidrule(lr){6-9}\cmidrule(lr){10-13}

& Cov & T1E & RMSE & RelB
& Cov & T1E & RMSE & RelB
& Cov & T1E & RMSE & RelB \\
\midrule

M1 & 0.858 & 0.138 & 0.302 & -0.003 & 0.880 & 0.118 & 0.190 & 0.006 & 0.893 & 0.110 & 0.150 & 0.002 \\
M2 & 0.986 & 0.013 & 0.319 & -0.005 & 0.909 & 0.091 & 0.202 & 0.003 & 0.912 & 0.087 & 0.159 & 0.003 \\
M3 & 0.836 & 0.173 & 0.300 & -0.005 & 0.812 & 0.182 & 0.191 & 0.004 & 0.807 & 0.194 & 0.152 & 0.000 \\
M4 & 0.874 & 0.131 & 0.319 & -0.004 & 0.840 & 0.156 & 0.199 & 0.003 & 0.826 & 0.173 & 0.159 & 0.000 \\
M5a & NA & NA & 0.319 & -0.004 & 0.826 & 0.174 & 0.226 & 0.002 & 0.881 & 0.110 & 0.189 & 0.005 \\
M5b & NA & NA & NA & NA & 0.955 & 0.039 & 0.226 & 0.002 & 0.954 & 0.041 & 0.189 & 0.005 \\
M6a & NA & NA & NA & NA & 0.955 & 0.039 & 0.226 & 0.002 & 0.954 & 0.041 & 0.189 & 0.005 \\
M6b & NA & NA & NA & NA & 0.948 & 0.048 & 0.199 & 0.003 & 0.953 & 0.046 & 0.159 & 0.000 \\
\bottomrule
\end{tabular}
}
\end{table}

\begin{table}[H]
\centering
\caption{Simulation results for CAC = 0.50, ICC = 0.10 (where $i$ = \#clusters, $j$ = \#periods).}
\label{tab:size_mode_between_and_within_cv1_cac050_icc010_period2}

\renewcommand{\arraystretch}{1.15}
\setlength{\tabcolsep}{4.5pt}

\resizebox{0.925\textwidth}{!}{%
\begin{tabular}{l cccc cccc cccc}
\toprule

\multicolumn{1}{c}{Model}
& \multicolumn{4}{c}{$(i,j)=(2,2)$}
& \multicolumn{4}{c}{$(i,j)=(4,2)$}
& \multicolumn{4}{c}{$(i,j)=(6,2)$} \\
\cmidrule(lr){2-5}\cmidrule(lr){6-9}\cmidrule(lr){10-13}

& Cov & T1E & RMSE & RelB
& Cov & T1E & RMSE & RelB
& Cov & T1E & RMSE & RelB \\
\midrule

M1 & 0.812 & 0.178 & 0.362 & -0.024 & 0.863 & 0.149 & 0.226 & 0.003 & 0.871 & 0.124 & 0.190 & 0.002 \\
M2 & 0.976 & 0.021 & 0.368 & -0.027 & 0.885 & 0.123 & 0.236 & 0.006 & 0.882 & 0.114 & 0.195 & 0.000 \\
M3 & 0.758 & 0.243 & 0.354 & -0.021 & 0.740 & 0.270 & 0.232 & 0.002 & 0.708 & 0.277 & 0.197 & 0.001 \\
M4 & 0.804 & 0.195 & 0.367 & -0.025 & 0.760 & 0.245 & 0.237 & 0.004 & 0.724 & 0.272 & 0.201 & 0.001 \\
M5a & NA & NA & 0.367 & -0.025 & 0.825 & 0.175 & 0.256 & 0.009 & 0.876 & 0.116 & 0.216 & 0.001 \\
M5b & NA & NA & NA & NA & 0.958 & 0.044 & 0.256 & 0.009 & 0.952 & 0.046 & 0.216 & 0.001 \\
M6a & NA & NA & NA & NA & 0.958 & 0.044 & 0.256 & 0.009 & 0.952 & 0.046 & 0.216 & 0.001 \\
M6b & NA & NA & NA & NA & 0.950 & 0.051 & 0.237 & 0.004 & 0.947 & 0.052 & 0.201 & 0.001 \\
\bottomrule
\end{tabular}
}
\end{table}

\begin{table}[H]
\centering
\caption{Simulation results for CAC = 0.70, ICC = 0.01 (where $i$ = \#clusters, $j$ = \#periods).}
\label{tab:size_mode_between_and_within_cv1_cac070_icc001_period2}

\renewcommand{\arraystretch}{1.15}
\setlength{\tabcolsep}{4.5pt}

\resizebox{0.925\textwidth}{!}{%
\begin{tabular}{l cccc cccc cccc}
\toprule

\multicolumn{1}{c}{Model}
& \multicolumn{4}{c}{$(i,j)=(2,2)$}
& \multicolumn{4}{c}{$(i,j)=(4,2)$}
& \multicolumn{4}{c}{$(i,j)=(6,2)$} \\
\cmidrule(lr){2-5}\cmidrule(lr){6-9}\cmidrule(lr){10-13}

& Cov & T1E & RMSE & RelB
& Cov & T1E & RMSE & RelB
& Cov & T1E & RMSE & RelB \\
\midrule

M1 & 0.930 & 0.076 & 0.240 & 0.011 & 0.933 & 0.064 & 0.141 & -0.006 & 0.939 & 0.057 & 0.107 & 0.001 \\
M2 & 0.994 & 0.007 & 0.280 & 0.007 & 0.956 & 0.049 & 0.167 & -0.011 & 0.949 & 0.045 & 0.122 & -0.001 \\
M3 & 0.920 & 0.087 & 0.246 & 0.005 & 0.919 & 0.077 & 0.140 & -0.006 & 0.924 & 0.073 & 0.107 & 0.000 \\
M4 & 0.939 & 0.067 & 0.279 & 0.007 & 0.939 & 0.068 & 0.159 & -0.011 & 0.930 & 0.063 & 0.120 & -0.001 \\
M5a & NA & NA & 0.279 & 0.007 & 0.833 & 0.171 & 0.203 & -0.011 & 0.888 & 0.109 & 0.162 & -0.001 \\
M5b & NA & NA & NA & NA & 0.960 & 0.042 & 0.203 & -0.011 & 0.958 & 0.035 & 0.162 & -0.001 \\
M6a & NA & NA & NA & NA & 0.960 & 0.042 & 0.203 & -0.011 & 0.958 & 0.035 & 0.162 & -0.001 \\
M6b & NA & NA & NA & NA & 0.954 & 0.047 & 0.159 & -0.011 & 0.957 & 0.033 & 0.120 & -0.001 \\

\bottomrule
\end{tabular}
}
\end{table}

\begin{table}[H]
\centering
\caption{Simulation results for CAC = 0.70, ICC = 0.05 (where $i$ = \#clusters, $j$ = \#periods).}
\label{tab:size_mode_between_and_within_cv1_cac070_icc005_period2}

\renewcommand{\arraystretch}{1.15}
\setlength{\tabcolsep}{4.5pt}

\resizebox{0.925\textwidth}{!}{%
\begin{tabular}{l cccc cccc cccc}
\toprule

\multicolumn{1}{c}{Model}
& \multicolumn{4}{c}{$(i,j)=(2,2)$}
& \multicolumn{4}{c}{$(i,j)=(4,2)$}
& \multicolumn{4}{c}{$(i,j)=(6,2)$} \\
\cmidrule(lr){2-5}\cmidrule(lr){6-9}\cmidrule(lr){10-13}

& Cov & T1E & RMSE & RelB
& Cov & T1E & RMSE & RelB
& Cov & T1E & RMSE & RelB \\
\midrule

M1 & 0.883 & 0.113 & 0.286 & -0.007 & 0.899 & 0.098 & 0.175 & 0.006 & 0.911 & 0.093 & 0.137 & 0.002 \\
M2 & 0.989 & 0.010 & 0.303 & -0.001 & 0.925 & 0.075 & 0.186 & 0.003 & 0.926 & 0.074 & 0.145 & 0.003 \\
M3 & 0.865 & 0.139 & 0.284 & -0.005 & 0.853 & 0.138 & 0.174 & 0.005 & 0.858 & 0.149 & 0.137 & 0.000 \\
M4 & 0.902 & 0.105 & 0.303 & -0.003 & 0.879 & 0.118 & 0.182 & 0.003 & 0.871 & 0.129 & 0.143 & 0.001 \\
M5a & NA & NA & 0.303 & -0.003 & 0.826 & 0.169 & 0.214 & 0.002 & 0.880 & 0.108 & 0.179 & 0.006 \\
M5b & NA & NA & NA & NA & 0.955 & 0.038 & 0.214 & 0.002 & 0.954 & 0.040 & 0.179 & 0.006 \\
M6a & NA & NA & NA & NA & 0.955 & 0.038 & 0.214 & 0.002 & 0.954 & 0.040 & 0.179 & 0.006 \\
M6b & NA & NA & NA & NA & 0.950 & 0.046 & 0.182 & 0.003 & 0.956 & 0.040 & 0.143 & 0.001 \\
\bottomrule
\end{tabular}
}
\end{table}

\begin{table}[H]
\centering
\caption{Simulation results for CAC = 0.70, ICC = 0.10 (where $i$ = \#clusters, $j$ = \#periods).}
\label{tab:size_mode_between_and_within_cv1_cac070_icc010_period2}

\renewcommand{\arraystretch}{1.15}
\setlength{\tabcolsep}{4.5pt}

\resizebox{0.925\textwidth}{!}{%
\begin{tabular}{l cccc cccc cccc}
\toprule

\multicolumn{1}{c}{Model}
& \multicolumn{4}{c}{$(i,j)=(2,2)$}
& \multicolumn{4}{c}{$(i,j)=(4,2)$}
& \multicolumn{4}{c}{$(i,j)=(6,2)$} \\
\cmidrule(lr){2-5}\cmidrule(lr){6-9}\cmidrule(lr){10-13}

& Cov & T1E & RMSE & RelB
& Cov & T1E & RMSE & RelB
& Cov & T1E & RMSE & RelB \\
\midrule

M1 & 0.850 & 0.141 & 0.330 & -0.017 & 0.888 & 0.120 & 0.199 & 0.002 & 0.888 & 0.105 & 0.166 & 0.000 \\
M2 & 0.986 & 0.018 & 0.334 & -0.021 & 0.909 & 0.100 & 0.209 & 0.004 & 0.898 & 0.099 & 0.171 & 0.000 \\
M3 & 0.811 & 0.190 & 0.321 & -0.018 & 0.811 & 0.195 & 0.201 & 0.001 & 0.778 & 0.207 & 0.169 & 0.001 \\
M4 & 0.857 & 0.149 & 0.334 & -0.022 & 0.826 & 0.171 & 0.206 & 0.003 & 0.792 & 0.201 & 0.172 & 0.001 \\
M5a & NA & NA & 0.334 & -0.022 & 0.826 & 0.178 & 0.234 & 0.009 & 0.878 & 0.113 & 0.197 & 0.001 \\
M5b & NA & NA & NA & NA & 0.956 & 0.043 & 0.234 & 0.009 & 0.950 & 0.043 & 0.197 & 0.001 \\
M6a & NA & NA & NA & NA & 0.956 & 0.043 & 0.234 & 0.009 & 0.950 & 0.043 & 0.197 & 0.001 \\
M6b & NA & NA & NA & NA & 0.951 & 0.049 & 0.206 & 0.003 & 0.950 & 0.048 & 0.172 & 0.001 \\
\bottomrule
\end{tabular}
}
\end{table}

\begin{table}[H]
\centering
\caption{Simulation results for CAC = 1.00, ICC = 0.01 (where $i$ = \#clusters, $j$ = \#periods).}
\label{tab:size_mode_between_and_within_cv1_cac100_icc001_period2}

\renewcommand{\arraystretch}{1.15}
\setlength{\tabcolsep}{4.5pt}

\resizebox{0.925\textwidth}{!}{%
\begin{tabular}{l cccc cccc cccc}
\toprule

\multicolumn{1}{c}{Model}
& \multicolumn{4}{c}{$(i,j)=(2,2)$}
& \multicolumn{4}{c}{$(i,j)=(4,2)$}
& \multicolumn{4}{c}{$(i,j)=(6,2)$} \\
\cmidrule(lr){2-5}\cmidrule(lr){6-9}\cmidrule(lr){10-13}

& Cov & T1E & RMSE & RelB
& Cov & T1E & RMSE & RelB
& Cov & T1E & RMSE & RelB \\
\midrule

M1 & 0.942 & 0.057 & 0.241 & -0.002 & 0.946 & 0.051 & 0.132 & 0.005 & 0.951 & 0.044 & 0.102 & -0.001 \\
M2 & 0.993 & 0.006 & 0.279 & 0.006 & 0.962 & 0.043 & 0.155 & 0.003 & 0.961 & 0.035 & 0.116 & -0.004 \\
M3 & 0.935 & 0.067 & 0.244 & 0.003 & 0.937 & 0.061 & 0.131 & 0.004 & 0.939 & 0.052 & 0.102 & -0.001 \\
M4 & 0.950 & 0.052 & 0.278 & 0.009 & 0.950 & 0.056 & 0.149 & 0.002 & 0.951 & 0.045 & 0.112 & -0.004 \\
M5a & NA & NA & 0.278 & 0.009 & 0.838 & 0.173 & 0.193 & -0.000 & 0.895 & 0.102 & 0.158 & -0.002 \\
M5b & NA & NA & NA & NA & 0.965 & 0.040 & 0.193 & -0.000 & 0.962 & 0.035 & 0.158 & -0.002 \\
M6a & NA & NA & NA & NA & 0.965 & 0.040 & 0.193 & -0.000 & 0.962 & 0.035 & 0.158 & -0.002 \\
M6b & NA & NA & NA & NA & 0.963 & 0.045 & 0.149 & 0.002 & 0.964 & 0.033 & 0.112 & -0.004 \\
\bottomrule
\end{tabular}
}
\end{table}

\begin{table}[H]
\centering
\caption{Simulation results for CAC = 1.00, ICC = 0.05 (where $i$ = \#clusters, $j$ = \#periods).}
\label{tab:size_mode_between_and_within_cv1_cac100_icc005_period2}

\renewcommand{\arraystretch}{1.15}
\setlength{\tabcolsep}{4.5pt}

\resizebox{0.925\textwidth}{!}{%
\begin{tabular}{l cccc cccc cccc}
\toprule

\multicolumn{1}{c}{Model}
& \multicolumn{4}{c}{$(i,j)=(2,2)$}
& \multicolumn{4}{c}{$(i,j)=(4,2)$}
& \multicolumn{4}{c}{$(i,j)=(6,2)$} \\
\cmidrule(lr){2-5}\cmidrule(lr){6-9}\cmidrule(lr){10-13}

& Cov & T1E & RMSE & RelB
& Cov & T1E & RMSE & RelB
& Cov & T1E & RMSE & RelB \\
\midrule

M1 & 0.924 & 0.068 & 0.265 & -0.010 & 0.940 & 0.055 & 0.150 & 0.007 & 0.953 & 0.049 & 0.111 & -0.000 \\
M2 & 0.997 & 0.006 & 0.278 & -0.009 & 0.961 & 0.037 & 0.160 & 0.009 & 0.963 & 0.034 & 0.116 & -0.001 \\
M3 & 0.919 & 0.079 & 0.259 & -0.010 & 0.926 & 0.068 & 0.146 & 0.007 & 0.941 & 0.063 & 0.109 & -0.001 \\
M4 & 0.950 & 0.050 & 0.278 & -0.011 & 0.949 & 0.050 & 0.152 & 0.007 & 0.950 & 0.047 & 0.113 & -0.001 \\
M5a & NA & NA & 0.278 & -0.011 & 0.833 & 0.170 & 0.196 & 0.010 & 0.890 & 0.104 & 0.158 & 0.003 \\
M5b & NA & NA & NA & NA & 0.957 & 0.041 & 0.196 & 0.010 & 0.961 & 0.039 & 0.158 & 0.003 \\
M6a & NA & NA & NA & NA & 0.957 & 0.041 & 0.196 & 0.010 & 0.961 & 0.039 & 0.158 & 0.003 \\
M6b & NA & NA & NA & NA & 0.952 & 0.045 & 0.152 & 0.007 & 0.964 & 0.034 & 0.113 & -0.001 \\
\bottomrule
\end{tabular}
}
\end{table}

\begin{table}[H]
\centering
\caption{Simulation results for CAC = 1.00, ICC = 0.10 (where $i$ = \#clusters, $j$ = \#periods).}
\label{tab:size_mode_between_and_within_cv1_cac100_icc010_period2}

\renewcommand{\arraystretch}{1.15}
\setlength{\tabcolsep}{4.5pt}

\resizebox{0.925\textwidth}{!}{%
\begin{tabular}{l cccc cccc cccc}
\toprule

\multicolumn{1}{c}{Model}
& \multicolumn{4}{c}{$(i,j)=(2,2)$}
& \multicolumn{4}{c}{$(i,j)=(4,2)$}
& \multicolumn{4}{c}{$(i,j)=(6,2)$} \\
\cmidrule(lr){2-5}\cmidrule(lr){6-9}\cmidrule(lr){10-13}

& Cov & T1E & RMSE & RelB
& Cov & T1E & RMSE & RelB
& Cov & T1E & RMSE & RelB \\
\midrule

M1 & 0.928 & 0.081 & 0.273 & -0.007 & 0.947 & 0.052 & 0.155 & -0.001 & 0.955 & 0.049 & 0.114 & 0.002 \\
M2 & 0.994 & 0.005 & 0.277 & -0.002 & 0.958 & 0.038 & 0.161 & -0.001 & 0.959 & 0.040 & 0.117 & 0.003 \\
M3 & 0.912 & 0.086 & 0.268 & 0.001 & 0.932 & 0.064 & 0.150 & -0.003 & 0.942 & 0.061 & 0.111 & 0.003 \\
M4 & 0.945 & 0.052 & 0.277 & -0.002 & 0.946 & 0.052 & 0.153 & -0.004 & 0.949 & 0.051 & 0.114 & 0.003 \\
M5a & NA & NA & 0.277 & -0.002 & 0.835 & 0.160 & 0.197 & 0.003 & 0.893 & 0.110 & 0.160 & 0.005 \\
M5b & NA & NA & NA & NA & 0.962 & 0.038 & 0.197 & 0.003 & 0.961 & 0.041 & 0.160 & 0.005 \\
M6a & NA & NA & NA & NA & 0.962 & 0.038 & 0.197 & 0.003 & 0.961 & 0.041 & 0.160 & 0.005 \\
M6b & NA & NA & NA & NA & 0.955 & 0.044 & 0.153 & -0.004 & 0.963 & 0.035 & 0.114 & 0.003 \\

\bottomrule
\end{tabular}
}
\end{table}

\subsection{Supplementary Results under Discrete Time Decay Correlation Structure (unequal cluster-period size)}

\begin{table}[H]
\centering
\caption{Simulation results for CAC = 0.50, ICC = 0.01 (where $i$ = \#clusters, $j$ = \#periods).}
\label{tab:decay_size_mode_between_and_within_cv1_cac050_icc001_period2}

\renewcommand{\arraystretch}{1.15}
\setlength{\tabcolsep}{4.5pt}

\resizebox{0.925\textwidth}{!}{%
\begin{tabular}{l cccc cccc cccc}
\toprule

\multicolumn{1}{c}{Model}
& \multicolumn{4}{c}{$(i,j)=(2,2)$}
& \multicolumn{4}{c}{$(i,j)=(4,2)$}
& \multicolumn{4}{c}{$(i,j)=(6,2)$} \\
\cmidrule(lr){2-5}\cmidrule(lr){6-9}\cmidrule(lr){10-13}

& Cov & T1E & RMSE & RelB
& Cov & T1E & RMSE & RelB
& Cov & T1E & RMSE & RelB \\
\midrule

M1 & 0.929 & 0.077 & 0.245 & 0.008 & 0.926 & 0.071 & 0.144 & -0.001 & 0.934 & 0.072 & 0.111 & 0.007 \\
M2 & 0.994 & 0.006 & 0.287 & 0.014 & 0.947 & 0.048 & 0.170 & -0.005 & 0.944 & 0.052 & 0.129 & 0.005 \\
M3 & 0.914 & 0.089 & 0.251 & 0.012 & 0.910 & 0.089 & 0.143 & -0.001 & 0.910 & 0.093 & 0.112 & 0.007 \\
M4 & 0.931 & 0.068 & 0.287 & 0.015 & 0.924 & 0.071 & 0.163 & -0.004 & 0.918 & 0.079 & 0.126 & 0.005 \\
M5a & NA & NA & 0.287 & 0.015 & 0.828 & 0.170 & 0.203 & -0.007 & 0.893 & 0.115 & 0.166 & 0.009 \\
M5b & NA & NA & NA & NA & 0.963 & 0.040 & 0.203 & -0.007 & 0.962 & 0.041 & 0.166 & 0.009 \\
M6a & NA & NA & NA & NA & 0.963 & 0.040 & 0.203 & -0.007 & 0.962 & 0.041 & 0.166 & 0.009 \\
M6b & NA & NA & NA & NA & 0.958 & 0.042 & 0.163 & -0.004 & 0.960 & 0.039 & 0.126 & 0.005 \\
\bottomrule
\end{tabular}
}
\end{table}

\begin{table}[H]
\centering
\caption{Simulation results for CAC = 0.50, ICC = 0.05 (where $i$ = \#clusters, $j$ = \#periods).}
\label{tab:decay_size_mode_between_and_within_cv1_cac050_icc005_period2}

\renewcommand{\arraystretch}{1.15}
\setlength{\tabcolsep}{4.5pt}

\resizebox{0.925\textwidth}{!}{%
\begin{tabular}{l cccc cccc cccc}
\toprule

\multicolumn{1}{c}{Model}
& \multicolumn{4}{c}{$(i,j)=(2,2)$}
& \multicolumn{4}{c}{$(i,j)=(4,2)$}
& \multicolumn{4}{c}{$(i,j)=(6,2)$} \\
\cmidrule(lr){2-5}\cmidrule(lr){6-9}\cmidrule(lr){10-13}

& Cov & T1E & RMSE & RelB
& Cov & T1E & RMSE & RelB
& Cov & T1E & RMSE & RelB \\
\midrule

M1 & 0.855 & 0.140 & 0.308 & 0.012 & 0.889 & 0.121 & 0.186 & -0.005 & 0.898 & 0.105 & 0.148 & 0.002 \\
M2 & 0.988 & 0.016 & 0.330 & 0.016 & 0.910 & 0.091 & 0.201 & -0.010 & 0.917 & 0.086 & 0.157 & 0.005 \\
M3 & 0.832 & 0.177 & 0.308 & 0.014 & 0.826 & 0.188 & 0.188 & -0.003 & 0.813 & 0.193 & 0.151 & 0.001 \\
M4 & 0.873 & 0.136 & 0.329 & 0.017 & 0.843 & 0.161 & 0.197 & -0.007 & 0.833 & 0.174 & 0.157 & 0.003 \\
M5a & NA & NA & 0.329 & 0.017 & 0.825 & 0.180 & 0.227 & -0.012 & 0.887 & 0.108 & 0.184 & 0.005 \\
M5b & NA & NA & NA & NA & 0.956 & 0.043 & 0.227 & -0.012 & 0.956 & 0.043 & 0.184 & 0.005 \\
M6a & NA & NA & NA & NA & 0.956 & 0.043 & 0.227 & -0.012 & 0.956 & 0.043 & 0.184 & 0.005 \\
M6b & NA & NA & NA & NA & 0.950 & 0.049 & 0.197 & -0.007 & 0.955 & 0.046 & 0.157 & 0.003 \\
\bottomrule
\end{tabular}
}
\end{table}

\begin{table}[H]
\centering
\caption{Simulation results for CAC = 0.50, ICC = 0.10 (where $i$ = \#clusters, $j$ = \#periods).}
\label{tab:decay_size_mode_between_and_within_cv1_cac050_icc010_period2}

\renewcommand{\arraystretch}{1.15}
\setlength{\tabcolsep}{4.5pt}

\resizebox{0.925\textwidth}{!}{%
\begin{tabular}{l cccc cccc cccc}
\toprule

\multicolumn{1}{c}{Model}
& \multicolumn{4}{c}{$(i,j)=(2,2)$}
& \multicolumn{4}{c}{$(i,j)=(4,2)$}
& \multicolumn{4}{c}{$(i,j)=(6,2)$} \\
\cmidrule(lr){2-5}\cmidrule(lr){6-9}\cmidrule(lr){10-13}

& Cov & T1E & RMSE & RelB
& Cov & T1E & RMSE & RelB
& Cov & T1E & RMSE & RelB \\
\midrule

M1 & 0.814 & 0.189 & 0.352 & -0.018 & 0.849 & 0.158 & 0.232 & -0.010 & 0.882 & 0.119 & 0.186 & -0.000 \\
M2 & 0.979 & 0.022 & 0.357 & -0.021 & 0.869 & 0.136 & 0.242 & -0.012 & 0.891 & 0.108 & 0.192 & 0.001 \\
M3 & 0.757 & 0.252 & 0.348 & -0.013 & 0.728 & 0.274 & 0.237 & -0.013 & 0.710 & 0.282 & 0.193 & 0.000 \\
M4 & 0.803 & 0.200 & 0.357 & -0.017 & 0.747 & 0.247 & 0.243 & -0.013 & 0.724 & 0.266 & 0.198 & 0.001 \\
M5a & NA & NA & 0.357 & -0.017 & 0.817 & 0.194 & 0.258 & -0.008 & 0.886 & 0.110 & 0.210 & -0.001 \\
M5b & NA & NA & NA & NA & 0.954 & 0.048 & 0.258 & -0.008 & 0.955 & 0.044 & 0.210 & -0.001 \\
M6a & NA & NA & NA & NA & 0.954 & 0.048 & 0.258 & -0.008 & 0.955 & 0.044 & 0.210 & -0.001 \\
M6b & NA & NA & NA & NA & 0.944 & 0.059 & 0.243 & -0.013 & 0.951 & 0.052 & 0.198 & 0.001 \\

\bottomrule
\end{tabular}
}
\end{table}

\begin{table}[H]
\centering
\caption{Simulation results for CAC = 0.70, ICC = 0.01 (where $i$ = \#clusters, $j$ = \#periods).}
\label{tab:decay_size_mode_between_and_within_cv1_cac070_icc001_period2}

\renewcommand{\arraystretch}{1.15}
\setlength{\tabcolsep}{4.5pt}

\resizebox{0.925\textwidth}{!}{%
\begin{tabular}{l cccc cccc cccc}
\toprule

\multicolumn{1}{c}{Model}
& \multicolumn{4}{c}{$(i,j)=(2,2)$}
& \multicolumn{4}{c}{$(i,j)=(4,2)$}
& \multicolumn{4}{c}{$(i,j)=(6,2)$} \\
\cmidrule(lr){2-5}\cmidrule(lr){6-9}\cmidrule(lr){10-13}

& Cov & T1E & RMSE & RelB
& Cov & T1E & RMSE & RelB
& Cov & T1E & RMSE & RelB \\
\midrule

M1 & 0.936 & 0.068 & 0.241 & 0.014 & 0.934 & 0.063 & 0.141 & -0.001 & 0.936 & 0.065 & 0.108 & 0.008 \\
M2 & 0.995 & 0.005 & 0.284 & 0.016 & 0.950 & 0.046 & 0.166 & -0.005 & 0.951 & 0.050 & 0.125 & 0.005 \\
M3 & 0.922 & 0.081 & 0.248 & 0.012 & 0.919 & 0.079 & 0.139 & -0.001 & 0.916 & 0.082 & 0.108 & 0.007 \\
M4 & 0.939 & 0.063 & 0.283 & 0.015 & 0.929 & 0.064 & 0.159 & -0.004 & 0.929 & 0.070 & 0.122 & 0.005 \\
M5a & NA & NA & 0.283 & 0.015 & 0.830 & 0.166 & 0.200 & -0.007 & 0.896 & 0.113 & 0.164 & 0.009 \\
M5b & NA & NA & NA & NA & 0.960 & 0.039 & 0.200 & -0.007 & 0.961 & 0.041 & 0.164 & 0.009 \\
M6a & NA & NA & NA & NA & 0.960 & 0.039 & 0.200 & -0.007 & 0.961 & 0.041 & 0.164 & 0.009 \\
M6b & NA & NA & NA & NA & 0.956 & 0.043 & 0.159 & -0.004 & 0.964 & 0.039 & 0.122 & 0.005 \\

\bottomrule
\end{tabular}
}
\end{table}

\begin{table}[H]
\centering
\caption{Simulation results for CAC = 0.70, ICC = 0.05 (where $i$ = \#clusters, $j$ = \#periods).}
\label{tab:decay_size_mode_between_and_within_cv1_cac070_icc005_period2}

\renewcommand{\arraystretch}{1.15}
\setlength{\tabcolsep}{4.5pt}

\resizebox{0.925\textwidth}{!}{%
\begin{tabular}{l cccc cccc cccc}
\toprule

\multicolumn{1}{c}{Model}
& \multicolumn{4}{c}{$(i,j)=(2,2)$}
& \multicolumn{4}{c}{$(i,j)=(4,2)$}
& \multicolumn{4}{c}{$(i,j)=(6,2)$} \\
\cmidrule(lr){2-5}\cmidrule(lr){6-9}\cmidrule(lr){10-13}

& Cov & T1E & RMSE & RelB
& Cov & T1E & RMSE & RelB
& Cov & T1E & RMSE & RelB \\
\midrule

M1 & 0.886 & 0.121 & 0.293 & 0.015 & 0.911 & 0.096 & 0.172 & -0.005 & 0.918 & 0.086 & 0.134 & 0.002 \\
M2 & 0.990 & 0.012 & 0.313 & 0.013 & 0.933 & 0.072 & 0.185 & -0.010 & 0.931 & 0.070 & 0.143 & 0.004 \\
M3 & 0.865 & 0.146 & 0.291 & 0.013 & 0.870 & 0.138 & 0.171 & -0.003 & 0.860 & 0.142 & 0.135 & 0.001 \\
M4 & 0.901 & 0.108 & 0.312 & 0.015 & 0.887 & 0.120 & 0.180 & -0.007 & 0.876 & 0.125 & 0.141 & 0.003 \\
M5a & NA & NA & 0.312 & 0.015 & 0.828 & 0.184 & 0.216 & -0.011 & 0.889 & 0.104 & 0.175 & 0.005 \\
M5b & NA & NA & NA & NA & 0.959 & 0.046 & 0.216 & -0.011 & 0.957 & 0.041 & 0.175 & 0.005 \\
M6a & NA & NA & NA & NA & 0.959 & 0.046 & 0.216 & -0.011 & 0.957 & 0.041 & 0.175 & 0.005 \\
M6b & NA & NA & NA & NA & 0.951 & 0.051 & 0.180 & -0.007 & 0.957 & 0.041 & 0.141 & 0.003 \\

\bottomrule
\end{tabular}
}
\end{table}

\begin{table}[H]
\centering
\caption{Simulation results for CAC = 0.70, ICC = 0.10 (where $i$ = \#clusters, $j$ = \#periods).}
\label{tab:decay_size_mode_between_and_within_cv1_cac070_icc010_period2}

\renewcommand{\arraystretch}{1.15}
\setlength{\tabcolsep}{4.5pt}

\resizebox{0.925\textwidth}{!}{%
\begin{tabular}{l cccc cccc cccc}
\toprule

\multicolumn{1}{c}{Model}
& \multicolumn{4}{c}{$(i,j)=(2,2)$}
& \multicolumn{4}{c}{$(i,j)=(4,2)$}
& \multicolumn{4}{c}{$(i,j)=(6,2)$} \\
\cmidrule(lr){2-5}\cmidrule(lr){6-9}\cmidrule(lr){10-13}

& Cov & T1E & RMSE & RelB
& Cov & T1E & RMSE & RelB
& Cov & T1E & RMSE & RelB \\
\midrule

M1 & 0.852 & 0.144 & 0.319 & -0.011 & 0.876 & 0.127 & 0.205 & -0.009 & 0.896 & 0.104 & 0.163 & -0.000 \\
M2 & 0.985 & 0.016 & 0.323 & -0.022 & 0.891 & 0.109 & 0.214 & -0.011 & 0.903 & 0.094 & 0.167 & 0.001 \\
M3 & 0.806 & 0.196 & 0.317 & -0.014 & 0.795 & 0.204 & 0.206 & -0.010 & 0.787 & 0.204 & 0.165 & 0.000 \\
M4 & 0.853 & 0.155 & 0.325 & -0.017 & 0.809 & 0.181 & 0.212 & -0.010 & 0.792 & 0.197 & 0.169 & 0.001 \\
M5a & NA & NA & 0.325 & -0.017 & 0.823 & 0.189 & 0.235 & -0.006 & 0.888 & 0.112 & 0.191 & -0.001 \\
M5b & NA & NA & NA & NA & 0.954 & 0.047 & 0.235 & -0.006 & 0.958 & 0.040 & 0.191 & -0.001 \\
M6a & NA & NA & NA & NA & 0.954 & 0.047 & 0.235 & -0.006 & 0.958 & 0.040 & 0.191 & -0.001 \\
M6b & NA & NA & NA & NA & 0.948 & 0.054 & 0.212 & -0.010 & 0.955 & 0.049 & 0.169 & 0.001 \\
\bottomrule
\end{tabular}
}
\end{table}

\begin{table}[H]
\centering
\caption{Simulation results for CAC = 1.00, ICC = 0.01 (where $i$ = \#clusters, $j$ = \#periods).}
\label{tab:decay_size_mode_between_and_within_cv1_cac100_icc001_period2}

\renewcommand{\arraystretch}{1.15}
\setlength{\tabcolsep}{4.5pt}

\resizebox{0.925\textwidth}{!}{%
\begin{tabular}{l cccc cccc cccc}
\toprule

\multicolumn{1}{c}{Model}
& \multicolumn{4}{c}{$(i,j)=(2,2)$}
& \multicolumn{4}{c}{$(i,j)=(4,2)$}
& \multicolumn{4}{c}{$(i,j)=(6,2)$} \\
\cmidrule(lr){2-5}\cmidrule(lr){6-9}\cmidrule(lr){10-13}

& Cov & T1E & RMSE & RelB
& Cov & T1E & RMSE & RelB
& Cov & T1E & RMSE & RelB \\
\midrule

M1 & 0.939 & 0.064 & 0.247 & 0.005 & 0.948 & 0.048 & 0.133 & 0.004 & 0.953 & 0.050 & 0.102 & -0.005 \\
M2 & 0.993 & 0.004 & 0.283 & 0.012 & 0.959 & 0.039 & 0.158 & 0.002 & 0.962 & 0.035 & 0.116 & -0.006 \\
M3 & 0.928 & 0.071 & 0.249 & 0.008 & 0.937 & 0.058 & 0.132 & 0.004 & 0.942 & 0.059 & 0.101 & -0.005 \\
M4 & 0.945 & 0.054 & 0.283 & 0.013 & 0.946 & 0.050 & 0.151 & 0.003 & 0.950 & 0.048 & 0.112 & -0.005 \\
M5a & NA & NA & 0.283 & 0.013 & 0.830 & 0.176 & 0.195 & -0.001 & 0.896 & 0.114 & 0.161 & -0.005 \\
M5b & NA & NA & NA & NA & 0.956 & 0.042 & 0.195 & -0.001 & 0.962 & 0.041 & 0.161 & -0.005 \\
M6a & NA & NA & NA & NA & 0.956 & 0.042 & 0.195 & -0.001 & 0.962 & 0.041 & 0.161 & -0.005 \\
M6b & NA & NA & NA & NA & 0.953 & 0.045 & 0.151 & 0.003 & 0.966 & 0.036 & 0.112 & -0.005 \\

\bottomrule
\end{tabular}
}
\end{table}

\begin{table}[H]
\centering
\caption{Simulation results for CAC = 1.00, ICC = 0.05 (where $i$ = \#clusters, $j$ = \#periods).}
\label{tab:decay_size_mode_between_and_within_cv1_cac100_icc005_period2}

\renewcommand{\arraystretch}{1.15}
\setlength{\tabcolsep}{4.5pt}

\resizebox{0.925\textwidth}{!}{%
\begin{tabular}{l cccc cccc cccc}
\toprule

\multicolumn{1}{c}{Model}
& \multicolumn{4}{c}{$(i,j)=(2,2)$}
& \multicolumn{4}{c}{$(i,j)=(4,2)$}
& \multicolumn{4}{c}{$(i,j)=(6,2)$} \\
\cmidrule(lr){2-5}\cmidrule(lr){6-9}\cmidrule(lr){10-13}

& Cov & T1E & RMSE & RelB
& Cov & T1E & RMSE & RelB
& Cov & T1E & RMSE & RelB \\
\midrule

M1 & 0.917 & 0.067 & 0.260 & 0.004 & 0.940 & 0.051 & 0.150 & 0.008 & 0.948 & 0.046 & 0.112 & -0.005 \\
M2 & 0.995 & 0.005 & 0.278 & 0.005 & 0.960 & 0.036 & 0.160 & 0.011 & 0.959 & 0.034 & 0.117 & -0.006 \\
M3 & 0.909 & 0.074 & 0.257 & 0.002 & 0.927 & 0.062 & 0.147 & 0.007 & 0.935 & 0.057 & 0.110 & -0.006 \\
M4 & 0.944 & 0.051 & 0.277 & 0.003 & 0.946 & 0.046 & 0.152 & 0.009 & 0.947 & 0.045 & 0.114 & -0.006 \\
M5a & NA & NA & 0.277 & 0.003 & 0.830 & 0.168 & 0.195 & 0.013 & 0.888 & 0.104 & 0.159 & -0.008 \\
M5b & NA & NA & NA & NA & 0.960 & 0.039 & 0.195 & 0.013 & 0.961 & 0.039 & 0.159 & -0.008 \\
M6a & NA & NA & NA & NA & 0.960 & 0.039 & 0.195 & 0.013 & 0.961 & 0.039 & 0.159 & -0.008 \\
M6b & NA & NA & NA & NA & 0.957 & 0.042 & 0.152 & 0.009 & 0.963 & 0.033 & 0.114 & -0.006 \\

\bottomrule
\end{tabular}
}
\end{table}

\begin{table}[H]
\centering
\caption{Simulation results for CAC = 1.00, ICC = 0.10 (where $i$ = \#clusters, $j$ = \#periods).}
\label{tab:decay_size_mode_between_and_within_cv1_cac100_icc010_period2}

\renewcommand{\arraystretch}{1.15}
\setlength{\tabcolsep}{4.5pt}

\resizebox{0.925\textwidth}{!}{%
\begin{tabular}{l cccc cccc cccc}
\toprule

\multicolumn{1}{c}{Model}
& \multicolumn{4}{c}{$(i,j)=(2,2)$}
& \multicolumn{4}{c}{$(i,j)=(4,2)$}
& \multicolumn{4}{c}{$(i,j)=(6,2)$} \\
\cmidrule(lr){2-5}\cmidrule(lr){6-9}\cmidrule(lr){10-13}

& Cov & T1E & RMSE & RelB
& Cov & T1E & RMSE & RelB
& Cov & T1E & RMSE & RelB \\
\midrule

M1 & 0.923 & 0.088 & 0.268 & -0.004 & 0.947 & 0.047 & 0.153 & -0.005 & 0.961 & 0.043 & 0.111 & 0.002 \\
M2 & 0.995 & 0.005 & 0.278 & 0.001 & 0.964 & 0.034 & 0.160 & -0.006 & 0.966 & 0.034 & 0.115 & 0.002 \\
M3 & 0.914 & 0.095 & 0.266 & -0.000 & 0.929 & 0.063 & 0.149 & -0.005 & 0.948 & 0.054 & 0.108 & 0.003 \\
M4 & 0.947 & 0.051 & 0.279 & 0.004 & 0.949 & 0.048 & 0.153 & -0.005 & 0.954 & 0.048 & 0.111 & 0.002 \\
M5a & NA & NA & 0.279 & 0.004 & 0.826 & 0.160 & 0.195 & -0.004 & 0.900 & 0.107 & 0.159 & 0.002 \\
M5b & NA & NA & NA & NA & 0.955 & 0.044 & 0.195 & -0.004 & 0.966 & 0.038 & 0.159 & 0.002 \\
M6a & NA & NA & NA & NA & 0.955 & 0.044 & 0.195 & -0.004 & 0.966 & 0.038 & 0.159 & 0.002 \\
M6b & NA & NA & NA & NA & 0.951 & 0.051 & 0.153 & -0.005 & 0.966 & 0.032 & 0.111 & 0.002 \\
\bottomrule
\end{tabular}
}
\end{table}

\subsection{Supplementary Results of convergence / usage rate in non-decay cases}
\begin{table}[H]
\centering
\caption{Usage and singular-fit rates for simulations with two clusters. Each cell reports usage / singular.}
\label{tab:convergence_nclus2}

\renewcommand{\arraystretch}{1.075}
\setlength{\tabcolsep}{3.0pt}

\resizebox{\textwidth}{!}{%
%
%
}
\end{table}

\begin{table}[H]
\centering
\caption{Usage and singular-fit rates for simulations with four clusters. Each cell reports usage / singular.}
\label{tab:convergence_nclus4}

\renewcommand{\arraystretch}{1.075}
\setlength{\tabcolsep}{3.0pt}

\resizebox{\textwidth}{!}{%
%
%
}
\end{table}

\begin{table}[H]
\centering
\caption{Usage and singular-fit rates for simulations with six clusters. Each cell reports usage / singular.}
\label{tab:convergence_nclus6}

\renewcommand{\arraystretch}{1.075}
\setlength{\tabcolsep}{3.0pt}

\resizebox{\textwidth}{!}{%
%
%
}
\end{table}

\subsection{Supplementary Results of convergence / usage rate in decay cases}
\begin{table}[H]
\centering
\caption{Usage and singular-fit rates under the discrete-time decay data-generating mechanism for simulations with two clusters. Each cell reports usage rate / singular rate.}
\label{tab:decay_convergence_nclus2}

\renewcommand{\arraystretch}{1.075}
\setlength{\tabcolsep}{3.0pt}

\resizebox{\textwidth}{!}{%
%
%
}

\end{table}

\begin{table}[H]
\centering
\caption{Usage and singular-fit rates under the discrete-time decay data-generating mechanism for simulations with four clusters. Each cell reports usage rate / singular rate.}
\label{tab:decay_convergence_nclus4}

\renewcommand{\arraystretch}{1.075}
\setlength{\tabcolsep}{3.0pt}

\resizebox{\textwidth}{!}{%
%
%
}
\end{table}

\begin{table}[htbp]
\centering
\caption{Usage and singular-fit rates under the discrete-time decay data-generating mechanism for simulations with six clusters. Each cell reports usage rate / singular rate.}
\label{tab:decay_convergence_nclus6}

\renewcommand{\arraystretch}{1.075}
\setlength{\tabcolsep}{3.0pt}

\resizebox{\textwidth}{!}{%
%
%
}
\end{table}

\end{document}